\pdfoutput=1

\documentclass[a4paper,12pt]{article}
\usepackage{fixltx2e}

\usepackage[margin=3cm,footskip=1cm]{geometry}

\usepackage[T1]{fontenc}
\usepackage[utf8]{inputenc}
\usepackage{color}
\usepackage{amsmath,amssymb,amsthm,bm,mathtools,xspace,booktabs,url}
\usepackage{graphicx,etoolbox}

\usepackage[shortlabels]{enumitem}
\setlist{
  listparindent=\parindent,
  parsep=0pt,
}

\SetEnumitemKey{:alph}{
  label=\textup{(\alph*)}
}
\SetEnumitemKey{:roman}{
  label=\textup{(\roman*)}
}

\setlist[enumerate]{
  :alph
}

\AtBeginEnvironment{thebibliography}{%
  \small
  \setlength\itemsep{0.75em plus 0.5em minus 0.75em}%
}

\usepackage{caption}
\usepackage[raggedright]{titlesec}

\numberwithin{equation}{section} 

\theoremstyle{plain} 
\newtheorem{theorem}{Theorem}[section]

\newtheorem{definition}[theorem]{Definition}

\theoremstyle{definition} 

\usepackage{authblk}

\makeatletter
\newcommand\CorrespondingAuthor[1]{%
  \begingroup%
  \def\@makefnmark{}%
  \footnotetext{Corresponding author: #1}%
  \endgroup%
}
\makeatother

\makeatletter
\renewenvironment{abstract}{%
  \small%
  \providecommand\keywords{%
    \par\medskip\noindent\textit{Keywords:}\xspace}%
  \begin{center}%
    \bfseries \abstractname\vspace{-.5em}\vspace{\z@}%
  \end{center}%
  \quote%
}{\endquote}
\makeatother

\usepackage[above,below,section]{placeins}

\usepackage[load-configurations=abbreviations]{siunitx}

\usepackage{lipsum}

\DeclarePairedDelimiter\abs\lvert\rvert
\DeclarePairedDelimiter\norm\lVert\rVert
\usepackage{xcolor}
\usepackage[draft]{fixme}
\fxsetup{ layout =marginnote, marginface=\scriptsize }

\newcommand\bx{\bm{x}}
\newcommand\by{\bm{y}}
\newcommand\bz{\bm{z}}
\newcommand\bX{\bm{X}}
\newcommand\bY{\bm{Y}}
\newcommand\bZ{\bm{Z}}

\newcommand\pp[1]{^{\smash{(#1)}\vphantom{gb}}}
\newcommand\ppm[1]{^{\smash{(#1)'}\vphantom{gb}}}

\DeclareMathOperator\spec{spec}
\DeclareMathOperator\DPP{DPP}

\newcommand\Rightarrowx{\quad\Rightarrow\quad}

\graphicspath{{.}{./code/pix-auto/}}

\newcommand\tsub[2]{_{\textup{#1},#2}}
\newcommand\sfrac[2]{#1/#2}

\begin{document}
\title{Determinantal point process models on the sphere}

\author[1]{Jesper M{\o}ller}

\author[1]{Morten Nielsen}

\author[2]{Emilio Porcu}

\author[1]{Ege Rubak}

\affil[1]{Department of Mathematical Sciences, Aalborg University,
  Denmark\authorcr jm@math.aau.dk. mnielsen@math.aau.dk,
  rubak@math.aau.dk}

\affil[2]{Department of Mathematics, University Federico Santa Maria,
  Chile\authorcr emilio.porcu@uv.cl}

\date{}

\maketitle

\begin{abstract}
  We consider determinantal point processes on the $d$-dimensional
  unit sphere~$\mathbb S^d$. These are finite point processes
  exhibiting repulsiveness and with moment properties determined by a
  certain determinant whose entries are specified by a so-called
  kernel which we assume is a complex covariance function defined on
  $\mathbb S^d\times\mathbb S^d$. We review the appealing properties
  of such processes, including their specific moment properties,
  density expressions and simulation procedures. Particularly, we
  characterize and construct isotropic DPPs models on~$\mathbb{S}^d$,
  where it becomes essential to specify the eigenvalues and
  eigenfunctions in a spectral representation for the kernel, and we
  figure out how repulsive isotropic DPPs can be.  Moreover, we
  discuss the shortcomings of adapting existing models for isotropic
  covariance functions and consider strategies for developing new
  models, including a useful spectral approach.

\end{abstract}

\section{Introduction}

Determinantal point processes (DPPs) are models for repulsiveness
(inhibition or regularity) between points in `space', where the two
most studied cases of `space' is a finite set or the $d$-dimensional
Euclidean space $\mathbb R^d$, though DPPs can be defined on fairly
general state spaces, cf.\ \cite{Hough:etal:09} and the references
therein.  DPPs are of interest because of their applications in
mathematical physics, combinatorics, random-matrix theory, machine
learning, and spatial statistics (see \cite{LMR1} and the references
therein). For DPPs on $\mathbb R^d$, rather flexible parametric models
can be constructed and likelihood and moment based inference
procedures apply, see \cite{LMR2,LMR1}.

This paper concerns models for DPPs defined on the $d$-dimensional
unit sphere $\mathbb{S}^d=\{\bx\in\mathbb R^{d+1}:\norm{\bx}=1\}$,
where $d\in\{1,2,\ldots\}$ and $\norm{\bx}$ denotes the usual
Euclidean distance, and where $d=1,2$ are the practically most
relevant cases. To the best of our knowledge, DPPs on $\mathbb{S}^d$
are largely unexplored in the literature, at least from a statistics
perspective.

\begin{figure}
  \includegraphics[width=\textwidth]{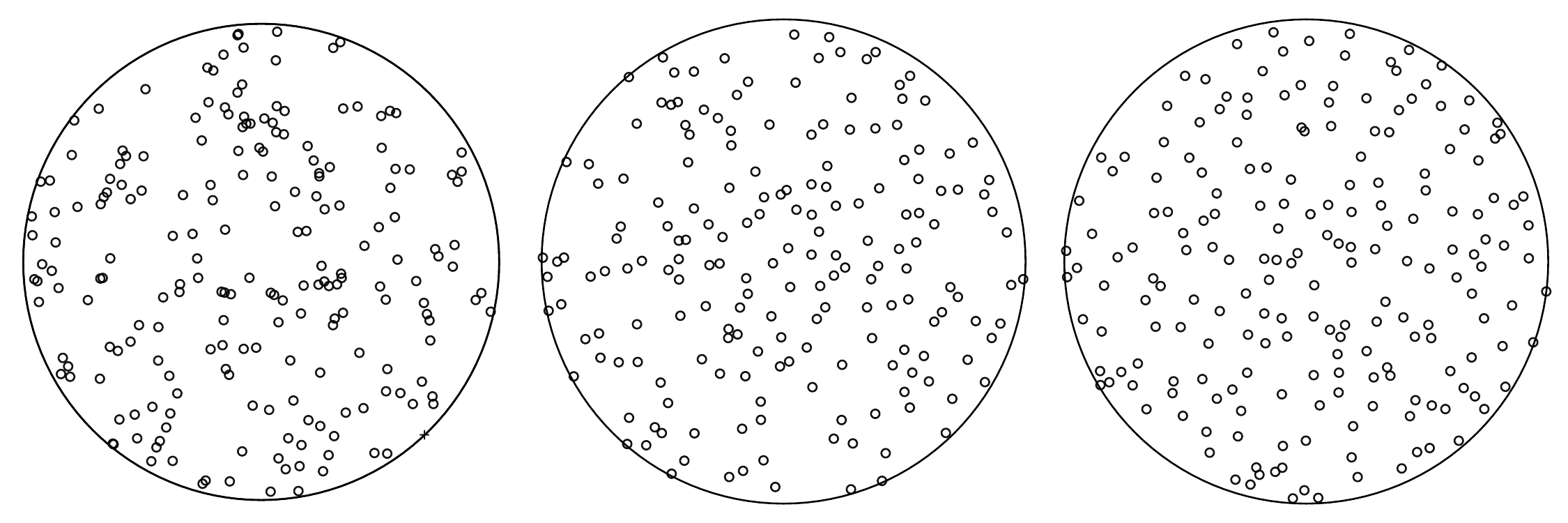}
  \caption{ Northern hemisphere of three spherical point patterns
    projected to the unit disc with an equal-area azimuthal
    projection.  Each pattern is a simulated realization of a
    determinantal point process on the sphere with mean number of
    points 400.  Left: Complete spatial randomness (Poisson process).
    Middle: Multiquadric model with $\tau=10$ and $\delta=0.74$ (see
    Section~\ref{s:negbino}).  Right: Most repulsive DPP (see
    Section~\ref{s:mostrep}).  }
  \label{fig:realizations}
\end{figure}

Section~\ref{sec:def} provides the precise definition of a DPP on
$\mathbb{S}^d$. Briefly, a DPP on $\mathbb{S}^d$ is a random finite
subset $\bX\subset\mathbb{S}^d$ whose distribution is specified by a
function $C:\mathbb{S}^d\times\mathbb{S}^d\mapsto\mathbb C$ called the
kernel, where $\mathbb C$ denotes the complex plane, and where $C$
determines the moment properties: the $n$th order joint intensity for
the DPP at pairwise distinct points
$\bx_1,\ldots,\bx_n\in\mathbb{S}^d$ agrees with the determinant of the
$n\times n$ matrix with $(i,j)$th entry $C(\bx_i,\bx_j)$. As in most
other work on DPPs, we restrict attention to the case where the kernel
is a complex covariance function.  We allow the kernel to be complex,
since this becomes convenient when considering simulation of DPPs, but
the kernel has to be real if it is isotropic (as argued in
Section~\ref{sec:construction}). As discussed in
Section~\ref{sec:def}, $C$ being a covariance function implies
repulsiveness, and a Poisson process is an extreme case of a DPP.  The
left panel in Figure~\ref{fig:realizations} shows a realization of a
Poisson process while the right panel shows a most repulsive DPP which
is another extreme case of a DPP studied in Section~\ref{s:mostrep}.
The middle panel shows a realization of a so-called multiquadric DPP
where the degree of repulsiveness is between these two extreme cases
(see Section~\ref{s:negbino}).

Section~\ref{sec:exist} discusses existence conditions for DPPs and
summarizes some of their appealing properties: their moment properties
and density expressions are known, and they can easily and quickly be
simulated.  These results depend heavily on a spectral representation
of the kernel based on Mercer's theorem. Thus finding the eigenvalues
and eigenfunctions becomes a central issue, and in contrast to DPPs on
$\mathbb R^d$ where approximations have to be used (see
\cite{LMR2,LMR1}), we are able to handle isotropic DPPs models on
$\mathbb{S}^d$, i.e., when the kernel is assumed to be isotropic.

Section~\ref{sec:construction}, which is our main section, therefore
focuses on characterizing and constructing DPPs models on
$\mathbb{S}^d$ with an isotropic kernel $C(\bx,\by)=C_0(s)$ where
$s=s(\bx,\by)$
is the geodesic (or orthodromic or great-circle) distance and where
$C_0$ is continuous and ensures that $C$ becomes a covariance
function. For recent efforts on covariance functions depending on the great circle distance, see \cite{Gneiting:2013,berg:porcu,porcu:bevilacqua:genton}.

As detailed in Section~\ref{sec:chariso}, $C_0$ has a
Schoenberg representation, i.e., it is a countable linear combination
of Gegenbauer polynomials (cosine functions if $d=1$; Legendre
polynomials if $d=2$) where the coefficients are nonnegative and
summable. We denote the sum of these coefficients by $\eta$, which
turns out to be the expected number of points in the DPP. In
particular we relate the Schoenberg representation to the Mercer
spectral representation from Section~\ref{sec:exist}, where the
eigenfunctions turn out to be complex spherical harmonic functions.
Thereby we can construct a number of tractable and flexible parametric
models for isotropic DPPs, by either specifying the kernel directly or
by using a spectral approach.  Furthermore, we notice the trade-off
between the degree of repulsiveness and how large $\eta$ can be, and
we figure out what the `most repulsive isotropic DPPs' are.  We also
discuss the shortcomings of adapting existing models for isotropic
covariance functions (as reviewed in \cite{Gneiting:2013}) when they
are used as kernels for DPPs on $\mathbb{S}^d$.

Section~\ref{s:final} contains our concluding remarks, including
future work on anisotropic DPPs on $\mathbb{S}^d$.

\section{Preliminaries}\label{sec:def}

Section~\ref{s:defdpp} defines and discusses what is meant by a DPP on
$\mathbb S^d$ in terms of joint intensities,
Section~\ref{s:assumptions} specifies certain regularity conditions,
and Section~\ref{sec:rem} discusses why there is repulsiveness in a
DPP.

\subsection{Definition of a DPP on the sphere}\label{s:defdpp}

We need to recall a few concepts and to introduce some notation.
 
For $d=1,2,\ldots$, let $\nu_d$ be the $d$-dimensional surface measure
on $\mathbb{S}^d\subset\mathbb R^{d+1}$, see e.g.\
\cite[Chapter~1]{MR3060033}. This can be defined recursively: For
$d=1$ and $\bx=(x_1,x_2)=(\cos\theta,\sin\theta)$ with
$0\le\theta<2\pi$, $\mathrm d\nu_1(\bx)=\mathrm d\theta$ is the usual
Lebesgue measure on $[0,2\pi)$. For $d\ge2$ and
$\bx=(\by\sin\vartheta,\cos\vartheta)$ with $\by\in\mathbb S^{d-1}$
and $\vartheta\in[0,\pi]$,
\begin{equation*}
  \mathrm d\nu_d(\bx)=\sin^{d-1}\vartheta\,
  \mathrm d\nu_{d-1}(\by)\,\mathrm
  d\vartheta.
\end{equation*}
In particular,
if $d=2$ and
$\bx=(x_1,x_2,x_3)=(\sin\vartheta\cos\varphi,\sin\vartheta\sin\varphi,\cos\vartheta)$
where $\vartheta\in[0,\pi]$ is the polar latitude and
$\varphi\in[0,2\pi)$ is the polar longitude, $\mathrm
d\nu_2(\bx)=\sin\vartheta\,\mathrm d\varphi\,\mathrm
d\vartheta$.
Note that $\mathbb{S}^d$ has surface measure
$\sigma_d=\nu_d(\mathbb{S}^d)=\frac{2\pi^{(d+1)/2}}{\Gamma((d+1)/2)}$
($\sigma_1=2\pi$, $\sigma_2=4\pi$).

Consider a finite point process on $\mathbb{S}^d$ with no multiple
points; we can view this as a random finite set
$\bX\subset\mathbb{S}^d$.  For $n=1,2,\ldots$, suppose $\bX$ has $n$th
order {\it joint intensity} $\rho^{(n)}$ with respect to the product
measure $\nu_d^{(n)}=\nu_d\otimes\cdots\otimes\nu_d$ ($n$ times), that
is, for any Borel function $h:(\mathbb{S}^d)^n\mapsto[0,\infty)$,
\begin{equation}\label{e:defrhon}
  \mathrm E\sum_{\bx_1,\ldots,\bx_n\in\bX}^{\not=}h(\bx_1,\ldots,\bx_n)=
  \int h(\bx_1,\ldots,\bx_n)
  \rho^{(n)}(\bx_1,\ldots,\bx_n)\,\mathrm
  d\nu_d^{(n)}(\bx_1,\ldots,\bx_n),
\end{equation}
where the expectation is with respect to the distribution of $\bX$ and $\not=$ over the
summation sign means that 
the sum is over all $\mathbf x_1\in\mathbf X,\ldots,\mathbf x_n\in\mathbf X$ such that $\mathbf x_1,\ldots,\mathbf x_n$ are pairwise different, so unless $\mathbf X$ contains at least $n$ points the sum is zero.
In particular, $\rho(\bx)=\rho\pp{1}(\bx)$ is the intensity function
(with respect to $\nu_d$).  Intuitively, if $\bx_1,\ldots,\bx_n$ are
pairwise distinct points on $\mathbb{S}^d$, then
$\rho\pp{n}(\bx_1,\ldots,\bx_n)\,\mathrm
d\nu\pp{n}_d(\bx_1,\ldots,\bx_n)$ is the probability that $\bX$ has a
point in each of $n$ infinitesimally small regions on $\mathbb{S}^d$
around $\bx_1,\ldots,\bx_n$ and of `sizes' $\mathrm
d\nu_d(\bx_1),\ldots,$ $\mathrm d\nu_d(\bx_n)$, respectively. Note
that $\rho\pp{n}$ is uniquely determined except on a
$\nu_d\pp{n}$-nullset.
\begin{definition}\label{d:1}
  Let $C:\mathbb{S}^d\times\mathbb{S}^d\mapsto\mathbb C$ be a mapping
  and $\bX\subset\mathbb S^d$ be a finite point process. We say that
  $\bX$ is a \emph{determinantal point process (DPP) on $\mathbb{S}^d$
    with kernel} $C$ and write $\bX\sim\DPP_d(C)$ if for all
  $n=1,2,\ldots$ and $\bx_1,\ldots,\bx_n\in\mathbb S^d$, $\bX$ has
  $n$th order joint intensity
  \begin{equation}\label{e:defdpp}
    \rho^{(n)}(\bx_1,\ldots,\bx_n)=
    \det\left(C(\bx_i,\bx_j)_{i,j=1,\ldots,n}\right), 
  \end{equation}
  where $\det\left(C(\bx_i,\bx_j)_{i,j=1,\ldots,n}\right)$ is the
  determinant of the $n\times n$ matrix with $(i,j)$th entry
  $C(\bx_i,\bx_j)$.
\end{definition}

Comments to Definition~\ref{d:1}:
\begin{enumerate}
\item If $\bX\sim\DPP_d(C)$, its {\it intensity function} is
  \begin{equation*}
    \rho(\bx)=C(\bx,\bx),\qquad \bx\in\mathbb S^d,
  \end{equation*}
  and the trace
  \begin{equation}\label{e:kappa11}
    \eta=\int C(\bx,\bx)\,\mathrm d\nu_d(\bx)
  \end{equation}
  is the expected number of points in $\bX$.
\item A Poisson process on $\mathbb S^d$ with a $\nu_d$-integrable
  intensity function $\rho$ is a DPP where the kernel on the diagonal
  agrees with $\rho$ and outside the diagonal is zero.  Another simple
  case is the restriction of the kernel of the Ginibre point process
  defined on the complex plane to $\mathbb S^1$, i.e., when
  \begin{equation}\label{e:ginibre}
    C(\bx_1,\bx_2)=\rho\exp\left[\exp\left\{i\left(\theta_1-\theta_2\right)\right\}\right],\qquad
    \bx_k=\exp\left(i\theta_k\right)\in\mathbb S^1,\ k=1,2.
  \end{equation} 
  (The Ginibre point process defined on the complex plane is a famous
  example of a DPP and it relates to random matrix theory, see e.g.\
  \cite{Ginibre:65,Hough:etal:09}; it is only considered in this paper
  for illustrative purposes.)
\item In accordance with our intuition, condition~\eqref{e:defdpp}
  implies that
  \begin{equation*}
    \rho^{(n+1)}(\bx_0,\ldots,\bx_{n})>0\Rightarrowx
    \rho^{(n)}(\bx_1,\ldots,\bx_n)>0
  \end{equation*}
  for any pairwise distinct points $\bx_0,\ldots,\bx_{n}\in\mathbb
  S^d$ with $n\ge1$. Condition~\eqref{e:defdpp} also implies that $C$
  must be positive semi-definite, since $\rho^{(n)}\ge0$.  In
  particular, by \eqref{e:defdpp}, $C$ is (strictly) positive definite
  if and only if
  \begin{equation}\label{e:pf}
    \parbox{0.7\linewidth}{
      \centering
      $\rho^{(n)}(\bx_1,\ldots,\bx_n)>0$ for $n=1,2,\ldots$
      and pairwise distinct points $\bx_1,\ldots,\bx_{n}\in\mathbb
      S^d$.}
  \end{equation} 
  The implication of the kernel being positive definite will be
  discussed several places further on.
\end{enumerate}

\subsection{Regularity conditions for the kernel}\label{s:assumptions}

Henceforth, as in most other publications on DPPs (defined on $\mathbb
R^d$ or some other state space), we assume that $C$ in
Definition~\ref{d:1}
\begin{itemize}
\item is a complex covariance function, i.e., $C$ is positive
  semi-definite and Hermitian,
\item is of finite trace class, i.e., $\eta<\infty$, cf.\
  \eqref{e:kappa11},
\item and $C\in L^2(\mathbb{S}^d\times\mathbb{S}^d,\nu_d^{(2)})$, the
  space of $\nu_d^{(2)}$ square integrable complex functions.
\end{itemize}
These regularity conditions become essential when we later work with
the spectral representation for $C$ and discuss various properties of
DPPs in Sections~\ref{sec:exist}--\ref{sec:construction}. Note that if
$C$ is continuous, then $\eta<\infty$ and $C\in
L^2(\mathbb{S}^d\times\mathbb{S}^d,\nu_d^{(2)})$.  For instance, the
regularity conditions are satisfied for the Ginibre DPP with kernel
\eqref{e:ginibre}.

\subsection{Repulsiveness}\label{sec:rem}

Since $C$ is a covariance function, condition~\eqref{e:defdpp} implies
that
\begin{equation}\label{e:compare}
  \rho^{(n)}(\bx_1,\ldots,\bx_n)\le\rho(\bx_1)\cdots\rho(\bx_n),
\end{equation}
with equality only if $\bX$ is a Poisson process with intensity
function $\rho$.  Therefore, since a Poisson process is the case of no
spatial interaction, a non-Poissonian DPP is repulsive.

For $\bx,\by\in\mathbb S^d$, let
\begin{equation*}
  R(\bx,\by)=\frac{C(\bx,\by)}{\sqrt{C(\bx,\bx)C(\by,\by)}}
\end{equation*}
be the correlation function corresponding to $C$ when
$\rho(\bx)\rho(\by)>0$, and define the {\it pair correlation function}
for $\bX$ by
\begin{equation}\label{e:g}
  g(\bx,\by)=\begin{cases}
    \frac{\rho^{(2)}(\bx,\by)}{\rho(\bx)\rho(\by)}=1-\abs{R(\bx,\by)}^2 
    & \text{if $\rho(\bx)\rho(\by)>0$.}
    \\
    0 & \text{otherwise.}
  \end{cases}
\end{equation}  
(This terminology for $g$ may be confusing, but it is adapted from
physics and is commonly used by spatial statisticians.) Note that
$g(\boldsymbol{x},\boldsymbol{x})=0$ and $g(\boldsymbol{x},\boldsymbol{y})\le1$ for all $\boldsymbol{x}\neq \boldsymbol{y}$, with equality only if
$\bX$ is a Poisson process, again showing that a DPP is repulsive.

\section{Existence, simulation, and density
  expressions}\label{sec:exist}

Section~\ref{s:mercerrep} recalls the Mercer (or spectral)
representation for a complex covariance function. This is used in
Section~\ref{sec:defbyints} to describe the existence condition and
some basic probabilistic properties of $\bX\sim\DPP_d(C)$,
including a density expression for $\bX$ which involves a certain
kernel $\tilde C$. Finally, Section~\ref{sec:defbydens} notices an
alternative way of specifying a DPP, namely in terms of the kernel
$\tilde C$.

\subsection{Mercer representation}\label{s:mercerrep}

We need to recall the spectral representation for a complex covariance
function $K:\mathbb{S}^d\times\mathbb{S}^d\mapsto\mathbb C$ which
could be the kernel $C$ of a DPP or the above-mentioned kernel $\tilde
C$.

Assume that $K$ is of finite trace class and is square integrable,
cf.\ Section~\ref{s:assumptions}.  Then, by Mercer's theorem (see
e.g.\ \cite[Section~98]{riesz:nagy}), ignoring a
$\nu_d\pp{2}$-nullset, we can assume that $K$ has spectral
representation
\begin{equation}\label{e:spectralrep}
  K(\bx,\by)=\sum_{n=1}^\infty\alpha_n Y_n(\bx)\overline{Y_n(\by)},
  \qquad \bx,\by\in\mathbb{S}^d,
\end{equation}
with
\begin{itemize}
\item absolute convergence of the series;
\item $Y_1,Y_2,\ldots$ being eigenfunctions which form an orthonormal
  basis for $L^2(\mathbb{S}^d,\nu_d)$, the space of $\nu_d$ square
  integrable complex functions;
\item the set of eigenvalues
  $\spec (K)=\{\alpha_1,\alpha_2,\ldots\}$ being unique, where
  each nonzero $\alpha_n$ is positive and has finite multiplicity, and
  the only possible accumulation point of the eigenvalues is 0;
\end{itemize}
see e.g.\ \cite[Lemma~4.2.2]{Hough:etal:09}.  If in addition $K$ is
continuous, then \eqref{e:spectralrep} converges uniformly and $Y_n$
is continuous if $\alpha_n\not= 0$.  We refer to \eqref{e:spectralrep}
as the {\it Mercer representation} of $K$ and call the eigenvalues for
the {\it Mercer coefficients}. Note that $\spec (K)$ is the
spectrum of $K$.

When $\bX\sim\DPP_d(C)$, we denote the Mercer coefficients of
$C$ by $\lambda_1,\lambda_2,\ldots$.  By \eqref{e:kappa11} and
\eqref{e:spectralrep}, the mean number of points in $\bX$ is then
\begin{equation}\label{e:kappa}
  \eta= 
  \sum_{n=1}^\infty\lambda_n.
\end{equation} 

For example, for the Ginibre DPP with kernel \eqref{e:ginibre},
\begin{equation}\label{e:CGinibre}
  C(\bx_1,\bx_2)=\rho\sum_{n=0}^\infty\frac{\exp\left\{i\,n(\theta_1-\theta_2)\right\}}{n!}\,.
\end{equation}
As the Fourier functions $\exp(in\theta)$, $n=0,1,\ldots$, are
orthogonal, the Mercer coefficients become $2\pi\rho/n!$,
$n=0,1,\ldots$.

\subsection{Results}\label{sec:defbyints}

Theorem~\ref{thm:dpp} below summarizes some fundamental probabilistic
properties for a DPP.  First we need a definition, noticing that in
the Mercer representation \eqref{e:spectralrep}, if
$\spec (K)\subseteq\{0,1\}$, then $K$ is a projection, since
$\int K(\bx,\bz)\overline{K(\bz,\by)}\,\mathrm
d\nu_d(\bz)=K(\bx,\by)$.

\begin{definition}\label{d:1a}
  If $\bX\sim\DPP_d(C)$ and $\spec (C)\subseteq\{0,1\}$,
  then $\bX$ is called a {\it determinantal projection point process}.
\end{definition}

\begin{theorem}\label{thm:dpp}
  Let $\bX\sim\DPP_d(C)$ where $C$ is a complex covariance
  function of finite trace class and $C\in L^2(\mathbb
  S^d\times\mathbb S^d,\nu_d^{(2)})$.
  \begin{enumerate}
  \item Existence of $\DPP_d(C)$ is equivalent to that
    \begin{equation}\label{e:existcond}
      \spec(C)\subset[0,1]
    \end{equation}
    and it is then unique.
  \item Suppose $\spec(C)\subset[0,1]$ and consider the
    Mercer representation
    \begin{equation}\label{e:CCC}
      C(\bx,\by)=\sum_{n=1}^\infty\lambda_n Y_n(\bx)\overline{Y_n(\by)},
      \qquad \bx,\by\in\mathbb{S}^d,
    \end{equation} 
    and let $B_1,B_2,\ldots$ be independent Bernoulli variables with
    means $\lambda_1,\lambda_2,\ldots$. Conditional on
    $B_1,B_2,\ldots$, let $\bY\sim\DPP_d(E)$ be the
    determinantal projection point process with kernel
    \begin{equation*}
      E(\bx,\by)=\sum_{n=1}^\infty B_n Y_n(\bx)\overline{Y_n(\by)},
      \qquad \bx,\by\in\mathbb{S}^d.
    \end{equation*}
    Then $\bX$ is distributed as $\bY$ (unconditionally on
    $B_1,B_2,\ldots$).
  \item Suppose $\spec (C)\subset\{0,1\}$.  Then the number
    of points in $\bX$ is constant and equal to $\eta=\int
    C(\bx,\bx)\,\mathrm d\nu_d(\bx)=\#\{n:\lambda_n=1\}$, and its
    density with respect to $\nu_d^{(\eta)}$ is
    \begin{equation}\label{e:density11}
      f_\eta(\{\bx_1,\ldots,\bx_n\})=
      \frac1{\eta!}\det\left(C(\bx_i,\bx_j)_{i,j=1,\ldots,\eta}\right),\qquad 
      \{\bx_1,\ldots,\bx_{\eta}\}\subset\mathbb S^d.
    \end{equation}
  \item Suppose $\spec (C)\subset[0,1)$. Let $\tilde C:
    \mathbb{S}^d\times\mathbb{S}^d\mapsto\mathbb C$ be the complex
    covariance function given by the Mercer representation sharing the
    same eigenfunctions as $C$ in \eqref{e:CCC} but with Mercer
    coefficients
    \begin{equation}\label{e:tildelambda}
      \tilde\lambda_n=\frac{\lambda_n}{1-\lambda_n},\qquad n=1,2,\ldots
    \end{equation} 
    Define
    \begin{equation*}
      D=\sum_{n=1}^\infty \log(1+\tilde\lambda_n).
    \end{equation*}
    Then $\DPP_d(C)$ is absolutely continuous with respect to
    the Poisson process on $\mathbb{S}^d$ with intensity measure
    $\nu_d$ and has density
    \begin{equation}\label{e:density}
      f(\{\bx_1,\ldots,\bx_n\})=\exp\left(\sigma_d-D\right)
      \det\bigl(\tilde C(\bx_i,\bx_j)_{i,j=1,\ldots,n}\bigr),
    \end{equation}
    for any finite point configuration
    $\{\bx_1,\ldots,\bx_n\}\subset\mathbb{S}^d$ ($n=0,1,\ldots$).
  \item Suppose $\spec (C)\subset[0,1)$ and $C$ is (strictly)
    positive definite. Then any finite subset of $\mathbb S^d$ is a
    feasible realization of $\bX$, i.e., $f(\{\bx_1,\ldots,\bx_n\})>0$
    for all $n=0,1,\ldots$ and pairwise distinct points
    $\bx_1,\ldots,\bx_n\in\mathbb S^d$.
  \end{enumerate}
\end{theorem}

Comments to Theorem~\ref{thm:dpp}:
\begin{enumerate}
\item This follows from \cite[Lemma~4.2.6 and
  Theorem~4.5.5]{Hough:etal:09}.  For example, for the Ginibre DPP on
  $\mathbb S^d$, it follows from \eqref{e:CGinibre} that $\eta\le1$,
  so this process is of very limited interest in practice.  We shall
  later discuss in more detail the implication of the condition
  \eqref{e:existcond} for how large the intensity and how repulsive a
  DPP can be.  Note that \eqref{e:existcond} and $C$ being of finite
  trace class imply that $C\in L^2(\mathbb S^d\times\mathbb
  S^d,\nu_d\pp{2})$.
\item\label{item:simulation} This fundamental result is due to
  \cite[Theorem~7]{Hough:etal:06} (see also
  \cite[Theorem~4.5.3]{Hough:etal:09}). It is used for simulating a
  realization of $\bX$ in a quick and exact way: Generate first the
  finitely many non-zero Bernoulli variables and second in a
  sequential way each of the $\sum_{n=1}^\infty B_n$ points in $\bY$,
  where a joint density similar to \eqref{e:density11} is used to
  specify the conditional distribution of a point in $\bY$ given the
  Bernoulli variables and the previously generated points in $\bY$.
  In \cite{LMR2,LMR1} the details for simulating a DPP defined on a
  $d$-dimensional compact subset of $\mathbb R^d$ are given, and with
  a change to spherical coordinates this procedure can immediately be
  modified to apply for a DPP on $\mathbb S^d$ (see Appendix~\ref{app:simulation} for an important technical detail which differs from $\mathbb R^d$).
\item This result for a determinantal projection point process is in
  line with (b).
\item For a proof of \eqref{e:density}, see e.g.\
  \cite[Theorem~1.5]{shirai:takahash:03}. If $n=0$ then we consider
  the empty point configuration $\emptyset$. Thus $\exp(-D)$ is the
  probability that $\bX=\emptyset$. Moreover, we have the following
  properties:
  \begin{itemize}
  \item $f$ is hereditary, i.e., for $n=1,2,\ldots$ and pairwise
    distinct points $\bx_0,\ldots,\allowbreak \bx_{n}\in\mathbb S^d$,
    \begin{equation}\label{e:hereditary}
      f(\{\bx_0,\ldots,\bx_{n}\})>0   \Rightarrowx
      f(\{\bx_1,\ldots,\bx_{n}\})>0.
    \end{equation}
    In other words, any subset of a feasible realization of $\bX$ is
    also feasible.
  \item If $\bZ$ denotes a unit rate Poisson process on $\mathbb S^d$,
    then
    \begin{equation}\label{e:rhoagain}
      \rho^{(n)}(\bx_1,\ldots,\bx_n)=\mathrm E f(\bZ\cup\{\bx_1,\ldots,\bx_n\})
    \end{equation}
    for any $n=1,2,\ldots$ and pairwise distinct points
    $\bx_1,\ldots,\bx_n\in\mathbb S^d$.
  \item $\tilde C$ is of finite trace class and $\tilde C\in
    L^2(\mathbb S^d\times\mathbb S^d,\nu_d\pp{2})$.
  \item There is a one-to-one correspondence between $C$ and $\tilde
    C$, where
    \begin{equation}\label{e:onetoone}
      \lambda_n=\frac{\tilde\lambda_n}{1+\tilde\lambda_n},\quad n=1,2,\ldots
    \end{equation}
  \end{itemize}
\item This follows by combining \eqref{e:pf}, \eqref{e:hereditary},
  and \eqref{e:rhoagain}.
\end{enumerate}

\subsection{Defining a DPP by its density}\label{sec:defbydens}
 
Alternatively, instead of starting by specifying the kernel $C$ of a
DPP on $\mathbb S^d$, if $\spec (C)\subset[0,1)$, the DPP may be
specified in terms of $\tilde C$ from the density expression
\eqref{e:density} by exploiting the one-to-one correspondence between
$C$ and $\tilde C$: First, we assume that $\tilde C:\mathbb
S^d\times\mathbb S^d\mapsto\mathbb C$ is a covariance function of
finite trace class and $\tilde C\in L^2(\mathbb S^d\times\mathbb
S^d,\nu_d\pp{2})$ (this is ensured if e.g.\ $\tilde C$ is
continuous). Second, we construct $C$ from the Mercer representation
of $\tilde C$, recalling that $\tilde C$ and $C$ share the same
eigenfunctions and that the Mercer coefficients for $C$ are given in
terms of those for $\tilde C$ by \eqref{e:onetoone}.  Indeed then
$\spec (C)\subset[0,1)$ and $\sum\lambda_n<\infty$, and so $\DPP_d(C)$
is well defined.

\section{Isotropic DPP models}\label{sec:construction}

Throughout this section we assume that $\bX\sim\DPP_d(C)$ where
$C$ is a continuous isotropic covariance function with
$\spec (C)\subset[0,1]$, cf.\ Theorem~\ref{thm:dpp}(a). Here
isotropy means that $C$ is invariant under the action of the
orthogonal group $O(d+1)$ on $\mathbb S^d$. In other words,
$C(\bx,\by)=C_0(s)$, where
\begin{equation*}\label{e:s}
  s=s(\bx,\by)=\arccos(\bx\cdot\by),\qquad\bx,\by\in\mathbb{S}^d,
\end{equation*}
is the geodesic (or orthodromic or great-circle) distance and $\cdot$
denotes the usual inner product on $\mathbb R^{d+1}$. Thus $C$ being
Hermitian means that $C_0$ is a real mapping: since
$C(\bx,\by)=C_0(s)=C(\by,\bx)$ and $C(\bx,\by)=\overline{C(\by,\bx)}$,
we see that $C(\by,\bx)=\overline{C(\by,\bx)}$ is real. Therefore,
$C_0$ is assumed to be a continuous mapping defined on $[0,\pi]$ such
that $C$ becomes positive semi-definite. Moreover, we follow
\cite{DJD_EP13} in calling $C_0:[0,\pi]\mapsto\mathbb R$ the {\it
  radial part} of $C$, and with little abuse of notation we write
$\bX\sim\DPP_d(C_0)$.

Note that some special cases are excluded: For a Poisson process with
constant intensity, $C$ is isotropic but not continuous. For the
Ginibre DPP, the kernel \eqref{e:ginibre} is a continuous covariance
function, but since the kernel is not real it is not isotropic (the
kernel is only invariant under rotations about the origin in the
complex plane).

Obviously, $\bX$ is invariant in distribution under the action of
$O(d+1)$ on $\mathbb S^d$. In particular, any point in $\bX$ is
uniformly distributed on $\mathbb S^d$.  Further, the intensity
\begin{equation*}
  \rho=C_0(0)
\end{equation*}
is constant and equal to the maximal value of $C_0$, while
\begin{equation*}
  \eta=\sigma_d C_0(0)
\end{equation*}
is the expected number of points in $\bX$.  Furthermore, assuming
$C_0(0)>0$ (otherwise $\bX=\emptyset$), the pair correlation function
is isotropic and given by
\begin{equation}\label{e:ggg}
  g(\bx,\by)=g_0(s)=1-R_0(s)^2,
\end{equation}
where
\begin{equation*}
  R_0(s)=C_0(s)/C_0(0)
\end{equation*} 
is (the radial part of) the correlation function associated to $C$.
Note that $g_0(0)=0$. For many examples of isotropic kernels for DPPs
(including those discussed later in this paper), $g_0$ will be a
non-decreasing function (one exception is the most repulsive DPP given
in Proposition~\ref{prop:mostrep} below).

In what follows, since we have two kinds of specifications for a DPP,
namely in terms of $C$ or $\tilde C$ (where in the latter case
$\spec (C)\subset[0,1)$, cf.\ Section~\ref{sec:defbydens}), let
us just consider a continuous isotropic covariance function $K:\mathbb
S^d\times\mathbb{S}^d\mapsto\mathbb R$.  Our aim is to construct
models for its radial part $K_0$ so that we can calculate the Mercer
coefficients for $K$ and the corresponding eigenfunctions and thereby
can use the results in Theorem~\ref{thm:dpp}.  As we shall see, the
case $d=1$ can be treated by basic Fourier calculus, while the case
$d\ge2$ is more complicated and involves surface spherical harmonic
functions and so-called Schoenberg representations.

In the sequel, without loss of generality, we assume $K_0(0)>0$ and
consider the normalized function $K_0(s)/K_0(0)$, i.e., the radial
part of the corresponding correlation
function. Section~\ref{sec:chariso} characterizes such functions so
that in Section~\ref{s:mostrep} we can quantify the degree of
repulsiveness in an isotropic DPP and in Section~\ref{s:parmodels} we
can construct examples of parametric models.

\subsection{Characterization of isotropic covariance functions on the
  sphere}\label{sec:chariso}

Gneiting \cite{Gneiting:2013} provided a detailed study of continuous
isotropic correlation functions on the sphere, with a view to
(Gaussian) random fields defined on $\mathbb S^d$.  This section
summarizes the results in \cite{Gneiting:2013} needed in this paper
and complement with results relevant for DPPs.

For $d=1,2,\ldots$, let $\Psi_d$ be the class of continuous functions
$\psi:[0,\pi]\mapsto\mathbb R$ such that $\psi(0)=1$ and the function
\begin{equation}\label{e:Rd}
  R_d(\bx,\by)=\psi(s), \qquad 
  \bx,\by\in\mathbb S^d,
\end{equation}
is positive semi-definite, where the notation stresses that $R_d$
depends on $d$ (i.e., $R_d$ is a continuous isotropic correlation
function defined on $\mathbb S^d\times\mathbb S^d$).
The classes $\Psi_d$ and $\Psi_{\infty}=\cap_{d=1}^{\infty}\Psi_d$ are
convex, closed under products, and closed under limits if the limit is
continuous, cf.\ \cite{MR0005922}.  Let $\Psi_d^+$ be the subclass of
those functions $\psi\in\Psi_d$ which are (strictly) positive
definite, and set $\Psi_{\infty}^+=\cap_{d=1}^{\infty}\Psi_d^+$,
$\Psi_d^- =\Psi_d\setminus \Psi_d^+$, and
$\Psi_{\infty}^-=\bigcap_{d=1}^{\infty}\Psi_d^-$. By
\cite[Corollary~1]{Gneiting:2013}, these classes are strictly
decreasing:
\begin{equation*}
  \Psi_1\supset\Psi_2\supset\cdots\supset\Psi_{\infty},\qquad 
  \Psi_1^+\supset\Psi_2^+\supset\cdots\supset\Psi_{\infty}^+,\qquad
  \Psi_1^-\supset\Psi_2^-\supset\cdots\supset\Psi_{\infty}^-,
\end{equation*}
and $\Psi_{\infty}=\Psi_{\infty}^+\cup\Psi_{\infty}^-$, where the
union is disjoint.

The following Theorem~\ref{thm:constlambda1} characterizes the class
$\Psi_d$ 
in terms of Gegenbauer polynomials and
so-called $d$-Schoenberg coefficients (this terminology is adapted
from \cite{DJD_EP13}). It also establishes the connection to the
Mercer representation of a continuous isotropic correlation function.

Recall that the Gegenbauer polynomial $\mathcal
C_\ell^{(\lambda)}:[-1,1]\mapsto\mathbb R$ of degree $\ell=0,1,\ldots$
is defined for $\lambda>0$ by the expansion
\begin{equation}\label{e:gegen}
  \frac{1}{(1+r^2-2r\cos s)^{\lambda}}=\sum_{\ell=0}^\infty r^\ell \mathcal
  C_\ell^{(\lambda)}(\cos s),\qquad -1<r<1,\qquad 0\le s\le\pi.
\end{equation}
We follow \cite{MR0005922} in defining
\begin{equation*}
  \mathcal C_\ell^{(0)}(\cos s)=\cos(\ell s),\qquad 0\le s\le\pi.
\end{equation*}
We have $C_\ell^{(0)}(1)=1$ and
$C_\ell^{(\frac{d-1}{2})}(1)=\binom{\ell+d-2}{\ell}$ for
$d=2,3,\ldots$.  Further, the Legendre polynomial of degree
$\ell=0,1,\ldots$ is by Rodrigues' formula given by
\begin{equation*}
  P_\ell(x)=\frac{1}{2^\ell\ell!}\frac{\mathrm d^\ell}{\mathrm
    dx^\ell}\{(x^2-1)^\ell\},\qquad -1<x<1,
\end{equation*}
and for $m=0,\ldots,\ell$, the associated Legendre functions
$P_\ell^{(m)}$ and $P_\ell^{(-m)}$ are given by
\begin{equation*}
  P_\ell^{(m)}(x) = (-1)^m\left(1-x^2\right)^{m/2}\frac{\mathrm d^m}
  {\mathrm dx^m}P_\ell(x),
  \qquad -1\le x\le 1,
\end{equation*}
and
\begin{equation*}
  P_\ell ^{(-m)} = (-1)^m \frac{(\ell-m)!}{(\ell+m)!} P_\ell ^{(m)}.
\end{equation*}
Note that $\mathcal C_\ell^{(\frac{1}{2})}=P_\ell$.  Furthermore, in
Theorem~\ref{thm:constlambda1}(b), $Y_{\ell,k,d}$ is a complex
spherical harmonic function and $\mathcal K_{\ell,d}$ is an index set
such that the functions $Y_{\ell,k,d}$ for $k\in\mathcal K_{\ell,d}$
and $\ell=0,1,\ldots$ are forming an orthonormal basis for
$L^2(\mathbb S^d,\nu_d)$.  Complex spherical harmonic functions are
constructed in e.g.\ \cite[Eq.~(2.5)]{Atsushi87}), but since their
general expression is rather complicated, we have chosen only to
specify these in Theorem~\ref{thm:constlambda1}(c)--(d) for the
practically most relevant cases $d=1,2$. Finally, letting
$m_{\ell,d}=\#\mathcal K_{\ell,d}$, then
\begin{equation*}
  m_{0,1}=1,\qquad m_{\ell,1}=2,\qquad\ell=1,2,\ldots,
\end{equation*}
and
\begin{equation}\label{e:bbbb}
  m_{\ell,d}=\frac{2\ell+d-1}{d-1}\binom{\ell+d-2}{\ell},\qquad
  \ell=0,1,\ldots,\qquad d=2,3,\ldots, 
\end{equation}
i.e., $m_{\ell,2}=2\ell+1$ for $\ell=0,1,\ldots$.

\begin{theorem}\label{thm:constlambda1}
  We have: 
  \begin{enumerate}
  \item $\psi\in\Psi_d$ if and only if $\psi$ is of the form
    \begin{equation}\label{e:K0}
      \psi(s) =\sum_{\ell=0}^{\infty} \beta_{\ell,d} \frac{
        {\cal C}_{\ell}^{(\frac{d-1}{2})}(\cos s)}{{\cal
          C}_{\ell}^{(\frac{d-1}{2})}(1)}, \qquad 0\le s\le\pi,
    \end{equation} 
    where the $d$-Schoenberg sequence $\beta_{0,d},\beta_{1,d},\ldots$
    is a probability mass function. Then, for
    $d=1$, $\psi\in\Psi_1^+$ if and only if for any two integers
    $0\le\ell<n$, there exists an integer $k\ge0$ such that
    $\beta_{\ell+kn,1}>0$. While, for
    $d\ge2$, $\psi\in\Psi_d^+$ if and only if the subsets of $d$-Schoenberg
    coefficients $\beta_{\ell,d}>0$ with an even respective odd index
    $\ell$ are infinite.
  \item For the correlation function $R_d$ in \eqref{e:Rd} with
    $\psi\in\Psi_d$ given by \eqref{e:K0}, the Mercer representation
    is
    \begin{equation}\label{e:Sch333}
      R_d(\bx,\by)=\sum_{\ell=0}^\infty\alpha_{\ell,d}
      \sum_{k\in\mathcal K_{\ell,d}}Y_{\ell,k,d}(\bx)\overline{Y_{\ell,k,d}}(\by),
    \end{equation}
    where the Mercer coefficient $\alpha_{\ell,d}$ is an eigenvalue of
    multiplicity $m_{\ell,d}$ and it is related to the $d$-Schoenberg
    coefficient $\beta_{\ell,d}$ by
    \begin{equation}\label{e:Sch3}
      \alpha_{\ell,d}=\sigma_{d}\frac{\beta_{\ell,d}}{m_{\ell,d}},\qquad\ell=0,1,\ldots
    \end{equation}
  \item Suppose $d=1$. Then the Schoenberg representation \eqref{e:K0}
    becomes
    \begin{equation}\label{e:Sch1111}
      \psi(s)=\sum_{\ell=0}^\infty\beta_{\ell,1}\cos(\ell s),\qquad 0 \le s \le \pi.
    \end{equation}
    Conversely,
    \begin{equation}\label{e:four}
      \beta_{0,1}=\frac{1}{\pi}\int_0^\pi\psi(s)\,\mathrm ds,\qquad 
      \beta_{\ell,1}=\frac{2}{\pi}\int_0^\pi\cos(\ell s)\psi(s)\,\mathrm
      ds,\qquad \ell=1,2,\ldots
    \end{equation}
    Moreover, for $R_1$ given by the Mercer representation
    \eqref{e:Sch333}, we have $\mathcal K_{0,1}=\{0\}$ and $\mathcal
    K_{\ell,1}=\{\pm1\}$ for $\ell>0$, and the eigenfunctions are the
    Fourier basis functions for $L^2(\mathbb S^1,\nu_1)$:
    \begin{equation*}
      Y_{\ell,k,1}(\theta)=
      \frac{\exp(ik\ell\theta)}{\sqrt{2\pi}},\qquad
      0\le\theta<2\pi,\qquad \ell=0,1,\ldots,\qquad k \in \mathcal K_{\ell,1}.
    \end{equation*}
  \item Suppose $d=2$. Then the Schoenberg representation
    \eqref{e:K0} becomes
    \begin{equation*}\label{eq:hov}
      \psi(s)=\sum_{\ell=0}^\infty 
      \beta_{\ell,2}P_{\ell}(\cos s),\qquad 0 \le s \le \pi.
    \end{equation*}
    Moreover, for $R_2$ given by the Mercer representation
    \eqref{e:Sch333}, $K_{\ell,2}=\{-\ell,\ldots,\ell\}$ and the
    eigenfunctions are the surface spherical harmonic functions given
    by
    \begin{equation}
      \label{eq:s_harmonics}
      \begin{multlined}
        Y_{\ell,k,2}( \vartheta , \varphi ) =
        \sqrt{
          \frac{2\ell+1}{4\pi}
          \,\frac{(\ell-k)!}{(\ell+k)!}
        }  \, P_\ell^{(k)} (
        \cos{\vartheta} ) \, e^{i k \varphi },\\
        (\vartheta,\varphi)\in[0,\pi]\times[0,2\pi),\quad 
        \ell=0,1,\ldots,\quad k \in \mathcal K_{\ell,2}.
      \end{multlined}
    \end{equation}
  \end{enumerate}
\end{theorem}

Comments to Theorem~\ref{thm:constlambda1}:
\begin{enumerate}
\item[(a)] Expression \eqref{e:K0} is a classical characterization
  result due to Schoenberg \cite{MR0005922}. For the other results in
  (a), see \cite[Theorem~1]{Gneiting:2013}.
\item[(b)] For $d=1$, \eqref{e:Sch333} is straightforwardly verified
  using basic Fourier calculus. For $d\ge2$, \eqref{e:Sch333} follows
  from \eqref{e:K0}, where ${\cal
    C}_{\ell}\pp{(d-1)/2}(1)=\binom{\ell+d-2}{\ell}$, and from
  the general addition formula for spherical harmonics (see e.g.\
  \cite[p.~10]{MR3060033}):
  \begin{equation}
    \label{e:additionformula}
    \sum_{k\in\mathcal
      K_{\ell,d}}Y_{\ell,k,d}(\bx)\overline{Y_{\ell,k,d}(\by)}=
    \frac{1}{\sigma_d}\frac{2\ell+d-1}{d-1}\mathcal
    C_\ell^{(\frac{d-1}{2})}(\cos s).
  \end{equation}
  When $R_d$ in \eqref{e:Sch333} is the correlation function $R_0$ for
  the kernel $C$ of the isotropic DPP $\bX$ with intensity $\rho$,
  note that
  \begin{equation}\label{e:nnn}
    \lambda_{\ell,k,d}=\lambda_{\ell,d}=\frac{\eta}{m_{\ell,d}}\beta_{\ell,d},\qquad
    k\in\mathcal K_{\ell,d},\qquad \ell=0,1,\ldots,  
  \end{equation}
  are the Mercer coefficients for $C$. Hence the range for the
  intensity is
  \begin{equation}\label{e:range}
    0<\rho\le\rho\tsub{max}{d},\qquad 
    \rho\tsub{max}{d}=
    \inf_{\ell:\,\beta_{\ell,d}>0}\frac{m_{\ell,d}}{\sigma_d\beta_{\ell,d}},
  \end{equation} 
  where $\rho\tsub{max}{d}$ is finite and as indicated in the notation
  may depend on the dimension $d$.

  In the special case where $\psi(s)$ is nonnegative, we prove in Appendix~\ref{app:infimum} that the infimum in \eqref{e:range} is attained at $\ell=0$ and consequently
  \begin{equation}\label{rho:max}
    \rho\tsub{max}{d}=
    \frac{m_{0,d}}{\sigma_d\beta_{0,d}}.
  \end{equation} 

  Note that the condition $\spec (C)\subset[0,1)$ is equivalent to
  $\rho<\rho\tsub{max}{d}$, and the kernel $\tilde C$ used in the
  density expression \eqref{e:density} is then as expected isotropic,
  with Mercer coefficients
  \begin{equation*}
    \tilde\lambda_{\ell,k,d}=\tilde\lambda_{\ell,d}=
    \frac{\eta\beta_{\ell,d}}{m_{\ell,d}-\eta\beta_{\ell,d}},\qquad
    k\in\mathcal K_{\ell,d},\qquad \ell=0,1,\ldots  .
  \end{equation*}
  This follows by combining \eqref{e:tildelambda}, \eqref{e:K0},
  \eqref{e:Sch3} and \eqref{e:nnn}.
\item[(c)] This follows straightforwardly from basic Fourier calculus.
\item[(d)] These results follow from \eqref{e:K0}--\eqref{e:Sch3}. (The
  reader mainly interested in the proof for case $d=2$ may consult
  \cite[Proposition 3.29]{marinucci:peccati:11} for the fact that the
  surface spherical harmonics given by \eqref{eq:s_harmonics}
  constitute an orthonormal basis for $L^2(\mathbb{S}^2,\nu_2)$, and
  then use \eqref{e:additionformula} where $\mathcal
  C_{\ell}\pp{(d-1)/2}=P_\ell$ for $d=2$.)
\end{enumerate}

For $d=1$, the inversion result \eqref{e:four} easily applies in many
cases.  When $d\ge2$, \cite[Corollary 2]{Gneiting:2013} (based on
\cite{MR0005922}) specifies the $d$-Schoenberg coefficients:
\begin{equation}\label{e:betainversion}
  \beta_{\ell,d}=\frac{2\ell+d-1}{2^{3-d}\pi}
  \frac{(\Gamma(\frac{d-1}{2}))^2}{\Gamma(d-1)}\int_0^\pi 
  \mathcal C_{\ell}^{(\frac{d-1}{2})}(\cos
  s)\sin^{d-1}(s)\psi(s)\,\mathrm ds,\qquad
  \ell=0,1,\ldots
\end{equation}
In general we find it hard to use this result, while it is much easier
first to find the so-called Schoenberg coefficients $\beta_{\ell}$
given in the following theorem and second to exploit their connection
to the $d$-Schoenberg coefficients (stated in \eqref{eq:coefficients}
below).

We need some further notation.  For non-negative integers $n$ and
$\ell$, define for $d=1$,
\begin{equation*}
  \gamma_{n,\ell}^{(1)}= 
  2^{-\ell}
  \left(2-\delta_{n,0}\delta_{\ell\!\!\!\!\pmod 2,0}\right) 
  \binom{\ell}{\frac{\ell-n}{2}},
\end{equation*}
where $\delta_{ij}$ is the Kronecker delta, and define for
$d=2,3,\ldots$,
\begin{equation*} 
  \gamma_{n,\ell}^{(d)}= 
  \frac{(2n+d-1)(\ell!)\Gamma(\frac{d-1}2)}{2^{\ell+1}\{(\frac{\ell-n}2)!\}\Gamma(\frac{\ell+n+d+1}2)} 
  \binom{n+d-2}{n}.
\end{equation*} 

\begin{theorem}\label{thm:constlambda2}
  We have:
  \begin{enumerate}
  \item $\psi\in\Psi_\infty$ if and only if $\psi$ is of the form
    \begin{equation}\label{e:K02}
      \psi(s) =\sum_{\ell=0}^{\infty} \beta_{\ell} \cos^{\ell}s,
      \qquad 0\le s\le\pi, 
    \end{equation} 
    where the Schoenberg sequence $\beta_{0},\beta_{1},\ldots$ is a
    probability mass function.  Moreover, $\psi\in\Psi_\infty^+$ if
    and only if the subsets of Schoenberg coefficients
    $\beta_{\ell}>0$ with an even respective odd index $\ell$ are
    infinite.
  \item For $\psi\in\Psi_\infty$ and $d=1,2,\ldots$, the
    $d$-Schoenberg sequence is given in terms of the Schoenberg
    coefficients by
    \begin{equation}\label{eq:coefficients}
      \beta_{n,d}=\sum_{\substack{\ell=n\\n-\ell\equiv 0\!\!\!\pmod 2}}^\infty\beta_{\ell} \gamma_{n,\ell}^{(d)},\qquad n=0,1,\ldots,
    \end{equation}
  \end{enumerate}
\end{theorem}

Comments to Theorem~\ref{thm:constlambda2}:

\begin{enumerate}
\item[(a)] Expression \eqref{e:K02} is a classical characterization
  result due to Schoenberg \cite{MR0005922}, while we refer to
  \cite[Theorem~1]{Gneiting:2013} for the remaining results.  It is
  useful to rewrite \eqref{e:K02} in terms of a probability generating
  function
  \begin{equation*}
    \varphi(x)=\sum_{\ell=0}^\infty
    x^{\ell}\beta_{\ell},\qquad -1\le x\le 1,
  \end{equation*}
  so that $\psi(s)=\varphi(\cos s)$.  Examples are given in
  Section~\ref{s:parmodels}.
\item[(b)] The relation \eqref{eq:coefficients} is verified in
    Appendix~\ref{app:proofconstlambda2}.

  Given the Schoenberg coefficients, \eqref{eq:coefficients} can be
  used to calculate the $d$-Schoen\-berg coefficients either exactly
  or approximately by truncating the sums in
  \eqref{eq:coefficients}. If there are only finitely many non-zero
  Schoenberg coefficients in \eqref{e:K02}, then there is only
  finitely many non-zero $d$-Schoenberg coefficients and the sums in
  \eqref{eq:coefficients} are finite.  Examples are given in
  Section~\ref{s:parmodels}.
\end{enumerate}

\subsection{Quantifying repulsiveness}\label{s:mostrep}

Consider again $\bX\sim\DPP_d(C_0)$ where $C_0=\rho R_0$,
$\rho>0$ is the intensity, and $R_0$ is the correlation function.  For
distinct points $\bx,\by\in\mathbb S^d$, recall that
\begin{equation*}
  \rho^2g_0(s(\bx,\by))\,\mathrm d\nu_d(\bx)\,\mathrm d\nu_d(\by)
\end{equation*}
is approximately the probability for $\bX$ having a point in each of
infinitesimally small regions on $\mathbb{S}^d$ around $\bx$ and $\by$
of `sizes' $\mathrm d\nu_d(\bx)$ and $\mathrm d\nu_d(\by)$,
respectively. Therefore, when seeing if a DPP is more repulsive than
another by comparing their pair correlation functions, we need to fix
the intensity. Naturally, we will say that
$\bX\pp{1}\sim\DPP_d(C_{0}\pp{1})$ is at least as repulsive than
$\bX\pp{2}\sim\DPP_d(C_{0}\pp{2})$ if they share the same intensity
and their pair correlation functions satisfy $g_0\pp{1}\le g_0\pp{2}$
(using an obvious notation). However, as pointed out in
\cite{LMR2,LMR1} such a simple comparison is not always possible.

Instead, following \cite{LMR2} (see also \cite{Lava-Biscio}), for an
arbitrary chosen point $\bx\in\mathbb S^d$, we quantify {\it global
  repulsiveness} of $\bX$ by
\begin{equation*}
  I(g_0)=\frac1{\sigma_d}
  \int_{\mathbb{S}^d}\left[1-g_0\left\{s(\bx,\by)\right\}\right]\,\mathrm
  d\nu_d(\by)=
  \frac1{\sigma_d}
  \int_{\mathbb{S}^d} R_0\left\{s(\bx,\by)\right\}^2\,\mathrm d\nu_d(\by)\,.
\end{equation*} 
Clearly, $I(g_0)$ does not depend on the choice of $\bx$, and $0\le
I(g_0)\le1$, where the lower bound is attained for a Poisson process
with constant intensity.  Furthermore, assuming $R_0(s)$ is twice
differentiable from the right at $s=0$, we quantify {\it local
  repulsiveness} of $\bX$ by the slope
\begin{equation*}
  g_0'(0)=-2R_0(0)R_0'(0)=-2R_0'(0)
\end{equation*} 
of the tangent line of the pair correlation function at $s=0$ and by
its curvature
\begin{equation*}
  c(g_0)=\frac{g_0''(0)}{\left\{1+g_0'^2(0)\right\}^{3/2}}
  =-2\frac{R_0'(0)^2+R_0''(0)}{\left\{1+4R_0'(0)^2\right\}^{3/2}}\,.
\end{equation*}
For many models we have $R_0'(0)=0$, and so $g_0'(0)=0$ and
$c(g_0)=g_0''(0)$.  In some cases, the derivative of $R_0(s)$ has a
singularity at $s=0$ (examples are given in Section~\ref{s:Matern});
then we define $g_0'(0)=\infty$.

\begin{definition}\label{d:repulsiveness}
  Suppose $\bX\pp{1}\sim\DPP_d(C_{0}\pp{1})$ and
  $\bX\pp{2}\sim\DPP_d(C_{0}\pp{2})$ share the same intensity $\rho>0$
  and have pair correlation functions $g_0\pp{1}$ and $g_0\pp{2}$,
  respectively.  We say that $\bX\pp{1}$ is at least as globally
  repulsive than $\bX\pp{2}$ if $I(g_0\pp{1})\ge I(g_0\pp{2})$. We say
  that $\bX\pp{1}$ is locally more repulsive than $\bX\pp{2}$ if
  either $g_0\pp{1}(s)$ and $g_0\pp{2}(s)$ are differentiable at $s=0$
  with $g\ppm{1}_0(0)< g\ppm{2}_0(0)$ or if $g_0\pp{1}(s)$ and
  $g_0\pp{2}(s)$ are twice differentiable at $s=0$ with
  $g\ppm{1}_0(0)= g\ppm{2}_0(0)$ and $c(g\pp{1}_0)< c(g\pp{2}_0)$.
\end{definition}

We think of the homogeneous Poisson process as the least globally and
locally repulsive DPP (for a given intensity), since its pair
correlation function satisfies $g_0(0)=0$ and $g_0(s)=1$ for $0\le
s\le\pi$ (this will be a limiting case in our examples to be discussed
in Section~\ref{s:parmodels}).  In what follows, we determine the most
globally and locally repulsive DPPs.

For $\eta=\sigma_d\rho>0$, let
$\bX^{(\eta)}\sim\DPP_d(C_0^{(\eta)})$ where $C_0^{(\eta)}$ has
Mercer coefficient $\lambda^{(\eta)}_{\ell,d}$ (of multiplicity
$m_{\ell,d}$) given by
\begin{equation}\label{e:g0kappa2}
  \lambda^{(\eta)}_{\ell,d}=1\ \mbox{if $\ell<n$,}
  \qquad \lambda^{(\eta)}_{n,d}=\frac{1}{m_{n,d}}
  \Bigl(\eta-\sum_{\ell=0}^{n-1} m_{\ell,d}\Bigr),\qquad
  \mbox{$\lambda^{(\eta)}_{\ell,d}=0$ if $\ell>n$,}
\end{equation}
where $n\ge0$ is the integer such that $\sum_{\ell=0}^{n-1}
m_{\ell,d}<\eta\le\sum_{\ell=0}^n m_{\ell,d}$, and where we set
$\sum_{\ell=0}^{-1} \cdots=0$. That is,
\begin{equation}\label{e:projmostrep}
  C_0^{(\eta)}(s)=\frac{1}{\sigma_d}\sum_{\ell=0}^{n-1} m_{\ell,d}
  \frac{\mathcal C_\ell^{(\frac{d-1}{2})}(\cos s)}{\mathcal
    C_\ell^{(\frac{d-1}{2})}(1)}
  +\frac{1}{\sigma_d}\Bigl(\eta-\sum_{\ell=0}^{n-1} 
    m_{\ell,d}\Bigr)
  \frac{\mathcal C_n^{(\frac{d-1}{2})}(\cos s)}{\mathcal
    C_n^{(\frac{d-1}{2})}(1)}\, .
\end{equation}
If $\eta=\sum_{\ell=0}^n m_{\ell,d}$, then $\bX\pp{\eta}$ is a
determinantal projection point process consisting of $\eta$ points,
cf.\ Theorem~\ref{thm:dpp}(b). If $\eta<\sum_{\ell=0}^n m_{\ell,d}$,
then $\bX\pp{\eta}$ is approximately a determinantal projection point
process and the number of points in $\bX\pp{\eta}$ is random with
values in $\{\sum_{\ell=0}^{n-1} m_{\ell,d},1+\sum_{\ell=0}^{n-1}
m_{\ell,d}, \ldots,\sum_{\ell=0}^{n} m_{\ell,d}\}$.  The following
proposition is verified in Appendix~\ref{app:proofmostrep}.

\begin{theorem}\label{prop:mostrep} 
  For a fixed value of the intensity $\rho>0$, we have:
  \begin{enumerate}
  \item $I(g_0)$ satisfies
    \begin{equation}\label{e:Ig0}
      \eta I(g_0)=1-\frac{1}{\eta}
      \sum_{\ell=0}^\infty m_{\ell,d} \lambda_{\ell,d}\left(1-\lambda_{\ell,d}\right),
    \end{equation}
    and so $\bX^{(\eta)}$ is a globally most repulsive isotropic DPP.
  \item If
    \begin{equation}\label{e:varcond}
      \sum_{\ell=1}^\infty\ell^2\beta_{\ell,d}<\infty,
    \end{equation}
    then $g_0'(0)=0$ and
    \begin{equation}\label{e:g0kappa}
      c(g_0)=g_0''(0)=\frac{2}{d}\sum_{\ell=1}^\infty
      \ell(\ell+d-1)\beta_{\ell,d} \, .
    \end{equation}
  \item $\bX^{(\eta)}$ is the unique locally most repulsive DPP
    among all isotropic DPPs satisfying \eqref{e:varcond}.
  \end{enumerate}
\end{theorem}

Comments to Theorem~\ref{prop:mostrep}:
\begin{enumerate}
\item It follows from \eqref{e:Ig0} that there may not be a unique
  globally most repulsive isotropic DPP, however, $\bX\pp{\eta}$
  appears to be the most natural one. For instance, if
  $\eta=\sum_{\ell=0}^n m_{\ell,d}$, there may exist another globally
  most repulsive determinantal projection point process with the
  non-zero Mercer coefficients specified by another finite index set
  $\mathcal L\subset\{0,1,\ldots\}$ than $\{0,\ldots,n\}$. In
  particular, for $d=1$ and $n>0$, there are infinitely many such
  index sets.

  By \eqref{e:Ig0},
  \begin{equation*}
    \eta I(g_0)\le1,
  \end{equation*}
  where the equality is obtained for $g_0=g_0\pp{\eta}$ when
  $\eta=\sum_{\ell=0}^n m_{\ell,d}$.  This inverse relationship
  between $\eta$ and
  $I(g_0)$ shows a \emph{trade-off between intensity and the degree of
    repulsiveness in a DPP}.
\item The variance condition \eqref{e:varcond} is sufficient to
  ensure twice differentiability from the right at 0 of $R_0$.  The
  condition is violated in the case of the exponential covariance
  function (the Mat\'{e}rn covariance function with $\nu=1/2$ and
  studied in Section~\ref{s:Matern}).
%
\item For simplicity, suppose that $\eta=\sum_{\ell=0}^n
  m_{\ell,d}$.

  Then $\bX\pp{\eta}$ is a determinantal projection point process
  consisting of $\eta$ points and with pair correlation function
  \begin{equation}\label{e:uniqueg}
    g_0^{(\eta)''}(0)=\frac{2\sum^n_{\ell=1}\ell(\ell+d-1)m_{\ell,d}}
    {d\sum_{\ell=0}^nm_{\ell,d}},
  \end{equation}
  cf.\ \eqref{e:g0kappa2} and \eqref{e:g0kappa}. Note that
  $g_0^{(\eta)''}(0)\sim n^2$ (here $f_1(\ell)\sim f_2(\ell)$ means
  that $c_d\le f_1(\ell)/f_2(\ell)\le C_d$ where $C_d\ge c_d$ are
  positive constants).

  For the practical important cases $d\le2$, we have that $\eta=2n+1$
  is odd if $d=1$, while $\eta=(n+1)^2$ is quadratic if $d=2$, cf.\
  \eqref{e:bbbb}. Furthermore, \eqref{e:projmostrep} simplifies to
  \begin{equation*}
    C_0^{(\eta)}(s)=\frac{1}{2\pi}\sum_{\ell=-n}^n \cos(\ell s)\qquad\mbox{if
      $d=1$},
  \end{equation*}
  and
  \begin{equation*}
    C_0^{(\eta)}(s)=\frac{1}{4\pi}\sum_{\ell=0}^n(2\ell+1)P_\ell(\cos s)
    \qquad\mbox{if $d=2$}.
  \end{equation*}
  Finally, a straightforward calculation shows that \eqref{e:uniqueg}
  becomes
  \begin{alignat}{2}
    \label{e:g01}
      g_0^{(\eta)''}(0)&=\tfrac{2}{3}n^2+\tfrac{2}{3}n&\qquad&\text{if $d=1$},
    \\
    \shortintertext{and}
    \label{e:g02}
      g_0^{(\eta)''}(0)&=\tfrac{1}{2}n^2+n&&\text{if $d=2$}.
    \end{alignat}
\end{enumerate}

\subsection{Parametric models}\label{s:parmodels}

In accordance with Theorem~\ref{prop:mostrep} we refer to
$\bX\pp{\eta}$ as `the most repulsive DPP' (when $\eta$ is
fixed). Ideally a parametric model class for the kernel of a DPP
should cover a wide range of repulsiveness, ranging from the most
repulsive DPP to the least repulsive DPP (the homogeneous Poisson
process).

This section considers parametric models for correlation functions
$\psi\in\Psi_d$ used to model
\begin{enumerate}[:roman]
\item either $C_0$ of the form
  \begin{equation}\label{e:(i)}
    C_0(s)=\rho\psi(s),\qquad 0<\rho\le\rho\tsub{max}{d},
  \end{equation}
  where $\rho\tsub{max}{d}<\infty$ is the upper bound on the
  intensity ensuring the existence of the DPP, cf.\ \eqref{e:range},
  noticing that $\rho\tsub{max}{d}$ depends on $\psi$;
\item or $\tilde C_0$ of the form
  \begin{equation}\label{e:(ii)}
    \tilde C_0(s)=\chi\psi(s),\qquad\chi>0,
  \end{equation}
  where $\tilde C_0$ is the radial part for $\tilde C$, and so the DPP
  is well defined for any positive value of the parameter $\chi$.
\end{enumerate}
In case (i) we need to determine the $d$-Schoenberg coefficients or at
least $\rho\tsub{max}{d}=\rho\tsub{max}{d}(\psi)$, and the
$d$-Schoenberg coefficients will also be needed when working with the
likelihood, cf.\ \eqref{e:density}.  In case (ii) we can immediately
work with the likelihood, while we need to calculate the
$d$-Schoenberg coefficients in order to find the intensity and the
pair correlation function.  In both cases, if we want to simulate from
the DPP, the $d$-Schoenberg coefficients have to be calculated.

In case (ii) with fixed $\psi$, the log-likelihood is simple to handle
with respect to the real parameter $\zeta=\ln\chi$: if
$\{x_1,\ldots,x_n\}$ is an observed point pattern with $n>0$ and
$\alpha_{\ell,d}$ is the $\ell$th Mercer coefficient for $\psi$, the
log-likelihood is
\begin{equation*}
  l(\zeta)=n\zeta+\ln\bigl[\det
  \{\psi(s(x_i,x_j))\}_{i,j=1,\ldots,n}\bigr]
  -\sum_{\ell=0}^\infty
  m_{\ell,d}\ln(1+\alpha_{\ell,d}\chi)\, ,
\end{equation*} 
cf.\ \eqref{e:density}. Hence the score function is
\begin{equation*}
  \frac{\mathrm dl(\zeta)}{\mathrm d\zeta}=
  n-\sum_{\ell=0}^\infty
  m_{\ell,d}\frac{\alpha_{\ell,d}\chi}{1+\alpha_{\ell,d}\chi}\, ,
\end{equation*}
and the observed information is
\begin{equation*}
  -\frac{\mathrm d^2l(\zeta)}{\mathrm d\zeta^2}=\sum_{\ell=0}^\infty m_{\ell,d}
  \frac{\alpha_{\ell,d}\chi}{\left(1+\alpha_{\ell,d}\chi\right)^2}\, ,
\end{equation*}
which is strictly positive (and agrees with the Fisher information).
Thus Newton-Raphson can be used for determining the maximum likelihood
estimate of $\chi$.

\subsubsection{Model strategies}

In general, when we start with a closed form expression for $\psi$,
the $d$-Schoenberg coefficients will be not be expressible on closed
form. Notable exceptions are a special case of the multiquadric family studied
in Section~\ref{s:negbino} and the spherical family and special cases
of the Askey and Wendland families (with $d\in\{1,3\}$) considered in
Section~\ref{s:wendland}.

Instead, if $\psi\in\Psi_\infty$, the Schoenberg coefficients may be expressed in closed form, making use of the identity in \eqref{eq:coefficients}. This is possible, for instance, for the multiquadric model. Other examples can be obtained by considering any probability mass system with a probability generating function being available in closed form. For instance, the binomial, Poisson, logathmic families can be used for such a setting. However,
in practice, if the sum in \eqref{eq:coefficients} is
infinite, a truncation will be needed so that approximate
$d$-Schoenberg coefficients are calculated (these will be smaller than
the true ones, so in case (i) above the approximation of the DPP is
still a well defined DPP). For instance, this is needed in case of a
Poisson distribution but not in case of a binomial distribution.

When the dimension $d$ is fixed, an alternative and as illustrated in
Section~\ref{s:flexible} often more flexible approach is to start by
modelling the Mercer coefficients for $\psi$. Then typically $\psi$
can only be expressed as an infinite sum (its $d$-Schoenberg
representation).

On the other hand, apart from the special cases considered in
Section~\ref{s:negbino}--\ref{s:wendland}, we have not been successful
in expressing Schoenberg coefficients on closed form for the `commonly
used' models for correlation functions, i.e., those listed in
\cite[Table~1]{Gneiting:2013}: the powered exponential, Mat{\'e}rn,
generalized Cauchy, etc. Moreover, these `commonly used' models seem
not very flexible for modelling repulsiveness. Section~\ref{s:Matern}
illustrates this in the case of the Mat{\'e}rn model.

\subsubsection{Multiquadric covariance functions}\label{s:negbino}

Let $p\in(0,1)$ and $\tau>0$ be the parameters of the negative
binomial distribution
\begin{equation*}
  \beta_\ell=\binom{\tau+\ell-1}{\ell}p^\ell(1-p)^\tau,\qquad
  \ell=0,1,\ldots
\end{equation*}
Then the corresponding Schoenberg representation reduces to
\begin{equation}\label{e:negb}
  \psi(s)=\left(\frac{1-p}{1-p\cos s}\right)^\tau,\qquad 0\le s\le\pi,
\end{equation}
where $\psi\in\Psi_\infty^+$, cf.\ Theorem~\ref{thm:constlambda2}(a).
This is the same as the multiquadric model in \cite{Gneiting:2013}
based on the reparametrization given by $p=\frac{2\delta}{1+\delta^2}$
with $\delta\in(0,1)$, since
\begin{equation}\label{e:multiquadricmodel}
  \psi(s)=\frac{(1-\delta)^{2\tau}}{(1+\delta^2-2\delta\cos
    s)^\tau}, \qquad 0\le s\le\pi.
\end{equation}
For $d=2$, we obtain for $\tau=\frac{1}{2}$ the inverse multiquadric
family and for $\tau=\frac{3}{2}$ the Poisson spline \cite{MR2610367}.
Furthermore, for $d=2$ we can solve \eqref{e:betainversion} explicitly and we have that the maximal 2-Schoenberg coefficient is
\begin{equation}\label{e:multiquadricbeta0}
    \beta_{0,2} =
    \begin{cases}
        \frac{(1-\delta)^{2\tau}}{4\delta(1-\tau)} \left( (1+\delta)^{2(1-\tau)} - (1-\delta)^{2(1-\tau)} \right) & \text{for } \tau \neq 1\\
        \frac{(1-\delta)^{2}}{2\delta} \log\left( \frac{1+\delta}{1-\delta} \right) & \text{for } \tau = 1
    \end{cases}
\end{equation}
which by \eqref{rho:max} gives us $\rho\tsub{max}{d}$.

Suppose $d\ge2$ and $\tau=\frac{d-1}{2}$. Then we derive directly from
\eqref{e:gegen} and \eqref{e:multiquadricmodel} that
\begin{equation}\label{e:betanegbin}
  \beta_{\ell,d}=\binom{\ell+d-2}{\ell}\delta^\ell(1-\delta)^{d-1}
  ,\qquad 
  \ell=0,1,\ldots,
\end{equation}
are the $d$-Schoenberg coefficients. Consider the case (i) where
$C_0=\rho\psi$, and let $\eta\tsub{max}{d}=\sigma_d\rho\tsub{max}{d}$ be
the maximal value of $\eta=\sigma_d\rho$ (the mean number of
points). Then
\begin{equation*}
  \eta\tsub{max}{d}= (1-\delta)^{1-d},
\end{equation*}
which is an increasing function of $\delta$, with range $(1,\infty)$,
and
\begin{equation*}
  \lambda_{\ell,d}=\frac{\eta}{\eta\tsub{max}{d}}
  \frac{d-1}{2\ell+d-1}\delta^\ell,\qquad
  \ell=0,1,\ldots
\end{equation*}
For any fixed value of $\eta>0$, as $\delta\rightarrow1$, we obtain
$\eta\tsub{max}{d}\rightarrow\infty$ and
$\lambda_{\ell,d}\rightarrow0$, corresponding to the Poisson process
with intensity $\rho$. On the other hand, the DPP is far from the most
repulsive DPP with the same value of $\eta$ unless $\eta$ is very
close to one: If $\eta=\eta\tsub{max}{d}>1$, then
\begin{equation*}
  \lambda_{\ell,d}=\frac{d-1}{2\ell+d-1}
  \left(1-\eta^{\frac{1}{1-d}}\right)^\ell,
\end{equation*} 
which is faster than algebraically decaying as a function of $\ell$
and is a strictly decreasing function of $\eta$ when $\ell>0$.  Also
the variance condition \eqref{e:varcond} is seen to be
satisfied. Hence, for $\eta=\eta\tsub{max}{d}>1$,
\begin{equation*}
  c(g_0)=g''(0)=
  \frac{2(d-1)\delta}{(1-\delta)^2}=
  2(d-1)\left(\eta^{\frac{2}{d-1}}-\eta^{\frac{1}{d-1}}\right)
\end{equation*}
which is an increasing function of $\eta$, with range $(0,\infty)$,
and $g''(0)$ is of order $\eta^{\frac{2}{d-1}}$. Thus we see again
that the DPP is far less repulsive than the most repulsive DPP with
the same value of $\eta$ (recalling that $g_0^{(\eta)''}(0)$ is of
order $\eta^{\sfrac{2}{d}}$).  To illustrate this, let $d=2$ and
$\eta=\eta\tsub{max}{2}=(1+n)^2$.  Then
\begin{equation*}
  g''(0)=2\left((n+1)^4-(n+1)^2\right),
\end{equation*}
which is of order $n^4$, while
$g_0^{((1+n)^2)''}(0)=\tfrac{1}{2}n^2+n$ is of order $n^2$ (the case
of the most repulsive DPP with $(1+n)^2$ points, cf.\ \eqref{e:g02}).

\begin{figure}
  \centering
  \includegraphics[width=.4\textwidth]{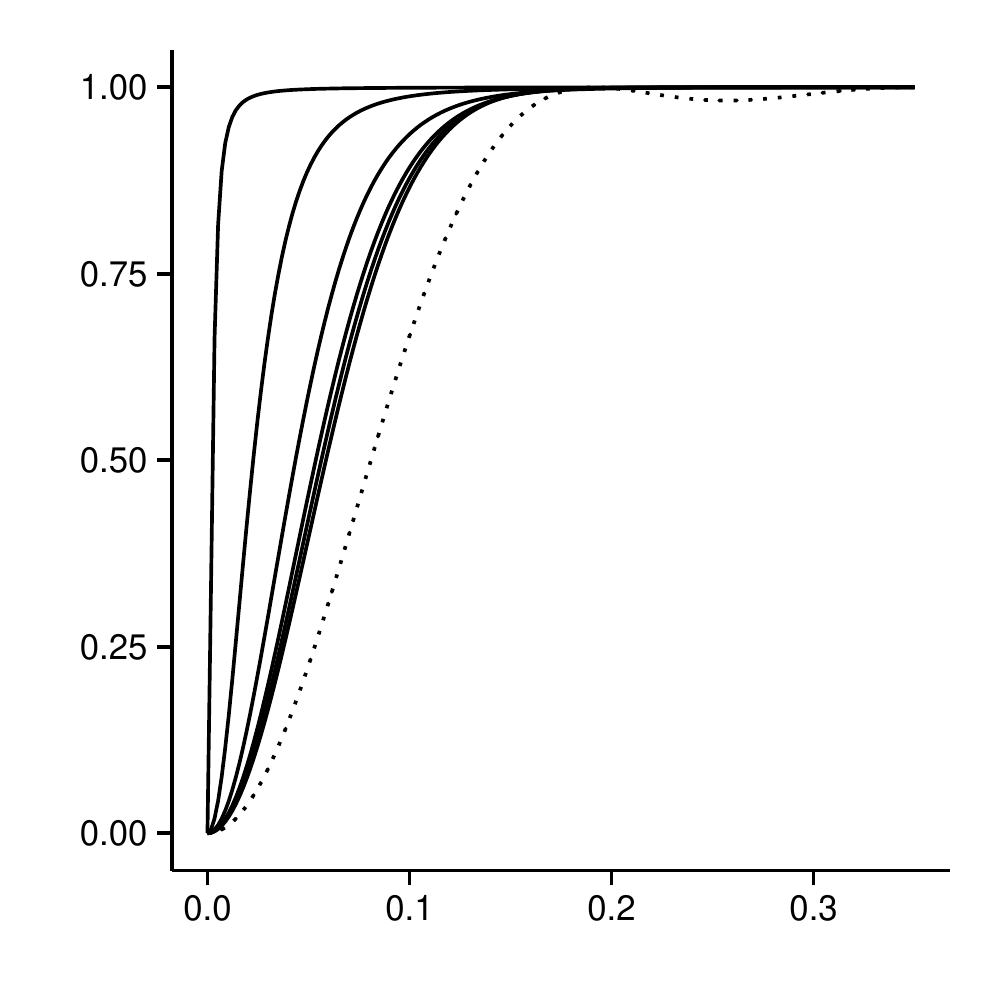}
  \caption{ Pair correlation functions for DPP models with $d=2$.
    Full lines from left to right correspond to multiquadric models
    with $\tau=1,\allowbreak 2,\allowbreak 5,\allowbreak
    10,\allowbreak 100$ and $\delta=0.97,\allowbreak 0.90,\allowbreak
    0.82,\allowbreak 0.74,\allowbreak 0.38$ chosen such that
    $\eta\tsub{max}{2}=400$.  The dotted line corresponds to the most
    repulsive DPP with $\eta=400$.  }
  \label{fig:MQpcf}
\end{figure}

In conclusion the inverse multiquadric model is not very flexible in
terms of the repulsiveness it can cover.  However, for other choices
of $\tau$ the situation appears to be much better.
Figure~\ref{fig:MQpcf} shows the pair correlation function for
different values of $\tau$ for $d=2$ when $\delta$ is chosen such that
$\eta\tsub{max}{2}=400$ together with the most repulsive DPP with
$\eta=400$.  The figure suggests that the models become more repulsive
when $\tau\to\infty$ with $\delta\to0$ appropriately chosen to keep
$\eta\tsub{max}{2}$ fixed, and based on the figure we conjecture that a limiting
model exists, but we have not been able to prove this.  The simulated
realization in the middle panel of Figure~\ref{fig:realizations} gives
the qualitative impression that the multiquadric model can obtain a
degree of repulsiveness that almost reaches the most repulsive DPP
though the corresponding pair correlation functions can easily be
distinguished in Figure~\ref{fig:MQpcf}.  To calculate $\eta\tsub{max}{2}$
we simply use \eqref{e:multiquadricbeta0} and \eqref{rho:max}.

\subsubsection{Spherical, Askey, and Wendland covariance
  functions}\label{s:wendland}

Table~\ref{table1} shows special cases of Askey's truncated power
function and $C^2$-Wendland and $C^4$-Wendland correlation functions
when $d\le3$ (here, for any real number, $x_+=x$ if $x\ge0$, and $x_+=0$ if
$x<0$). Note that a scale parameter $c$ can be included, where for the $C^2$ and $C^4$ Wendland correlation functions, $c \in (0,2\pi]$
and $c$ defines the compact support of these correlation functions, and where for the 
Askey's truncated power
function, $c>0$.
Notice that these correlation functions are of class
$\Psi_3^+$, they are compactly supported for $c<\pi$, and compared to
the Askey and Wendland correlation functions in
\cite[Table~1]{Gneiting:2013} they are the most repulsive cases.
Appendix~\ref{app:Askey} describes how the one-Schoenberg coefficients listed in
Table~\ref{table1} can be derived. The one-Schoenberg coefficients can
then be used to obtain the 3-Schoenberg coefficients, cf.\
\cite[Corollary~3]{Gneiting:2013}. Moreover, the spherical correlation
function
\begin{equation*}
  \psi(s)=\left(1+\frac{s}{2c}\right)\left(1-\frac{s}{c}\right)_+^2,
  \qquad 0\le s\le\pi,
\end{equation*}
is of class $\Psi_3^+$, and the proof of \cite[Lemma~2]{Gneiting:2013}
specifies its one-Schoenberg coefficients and hence the 3-Schoenberg
coefficients can also be calculated. Plots (omitted here) of the
corresponding Mercer coefficients for $d\in\{1,3\}$ show that DPPs
with the kernel specified by the Askey, $C^2$-Wendland,
$C^4$-Wendland, or spherical correlation function are very far from
the most repulsive case.

\begin{table}[htp]
  \caption{Special cases of Askey's truncated power
    function and $C^2$-Wendland and $C^4$-Wendland correlation
    functions      when       $d\le3$ and $c=1$, where the two last
    columns specify  the corresponding one-Schoenberg
    coefficients. For a general value of a scale parameter $c>0$ (with $c\le2\pi$ in case of the Wendland functions), 
    in the expressions for $\psi$, $s$ should be replaced
    by $s/c$, while in the expressions for
    $\beta_{\ell,1}$, $\ell$ should be replaced
    by $c\ell$ and the one-Schoenberg
    coefficient should be multiplied by $c$.}
    \label{table1}

  \centering
  \small
  \begin{tabular}{l l l l}
    \toprule
    & $\psi$ & $\beta_{\ell,1}$ ($\ell=1,2,\ldots$)& $\beta_{0,1}$
    \\
    \midrule
    Askey & $(1-s)_{+}^{3}$ & $\frac{6 (\ell^2+2 \cos
        (\ell)-2)}{\pi \ell^4}$ & $\frac{1}{4 \pi}$ 
    \\
    \addlinespace
    $C^2$-Wendland & $ (1-s)_+^4 (4 s+1)$ & $\frac{240
      (\ell^2+\ell \sin (\ell)+4 \cos (\ell)-4)}{\pi
      \ell^6}$ & $\frac{1}{6 \pi}$ 
    \\
    \addlinespace
    $C^4$-Wendland & $\frac{1}{3} (1-s)_+^6 (s(35 s+18)+3)$ &
    $\frac{8960 (-4 \ell (\ell^2-18)+3
        (\ell^2-35) \sin (\ell)+33 \ell \cos
        (\ell))}{\pi  \ell^9}$ & $\frac{4}{27 \pi}$ 
    \\
    \bottomrule
  \end{tabular}
\end{table}

\subsubsection{A flexible spectral model}\label{s:flexible}

Suppose that the kernel of the DPP has Mercer coefficients
\begin{equation*}\label{e:powerexp}
  \lambda_{\ell,d}=\frac{1}{1+\beta\exp\left((\ell/\alpha)^\kappa\right)}
  \,,\qquad \ell=0,1,\ldots,
\end{equation*}
where $\alpha>0$, $\beta>0$, and $\kappa>0$ are parameters. Since all
$\lambda_{\ell,d}\in(0,1)$, the DPP is well defined and has a density
specified by \eqref{e:density}. Since its kernel is positive definite,
all finite subsets of $\mathbb S^d$ are feasible realizations of the
DPP. The mean number of points $\eta$ may be evaluated by numerical
methods.

The most repulsive DPP is a limiting case: For any
$n\in\{0,1,\ldots\}$, let $\eta_0=\sum_{\ell=0}^n m_{\ell,d}$,
$\alpha=n$, and $\beta=1/(n\kappa)$. Then, as
$\kappa\rightarrow\infty$,
$\beta\exp\left((\ell/\alpha)^\kappa\right)$ converges to 0 for
$\ell\le n$ and to $\infty$ for $\ell> n$. Thus
$\lambda_{\ell,d}\rightarrow 1$ for $\ell\le n$,
$\lambda_{\ell,d}\rightarrow 0$ for $\ell> n$, and
$\eta\rightarrow\eta_0$. This limiting case corresponds to the case of
$\bX^{(\eta_0)}$, the most repulsive DPP consisting of $\eta_0$
points.

Also the homogeneous Poisson process is a limiting case: Note that for
fixed $d\geq 1$, the multiplicities $m_{\ell,d}$ given by
\eqref{e:bbbb} satisfy the asymptotic estimate $m_{l,d}\sim
(1+\ell)^{d-1}$ as $\ell\rightarrow \infty$ (again $f_1(\ell)\sim
f_2(\ell)$ means that $c_d\le f_1(\ell)/f_2(\ell)\le C_d$ where
$C_d\ge c_d$ are positive constants). Hence,
\begin{equation*}
  \eta=\sum_{\ell=0}^\infty m_{\ell,d} \lambda_{\ell,d}\sim 
  \sum_{\ell=0}^\infty (1+\ell)^{d-1} \frac{1}{1+\beta\exp((\ell/\alpha)^\kappa)}.
\end{equation*}
Now, for a given value $\eta=\eta_0>0$, put $\kappa=d$ and
$\alpha=(\eta_0\beta)^{1/d}$. Then, for sufficiently large $\beta$,
\begin{equation*}
  \sum_{\ell=0}^\infty (1+\ell)^{d-1} \frac{1}{1+\beta\exp((\ell/\alpha)^d)}\sim\eta_0.
\end{equation*}
Consequently, as $\beta\rightarrow \infty$, we obtain $\eta\sim
\eta_0$, while
\begin{equation*}
  \lambda_{\ell,d}\rightarrow 0,\qquad \ell=0,1,2\ldots
\end{equation*}

\begin{figure}
  \includegraphics[width=\textwidth]{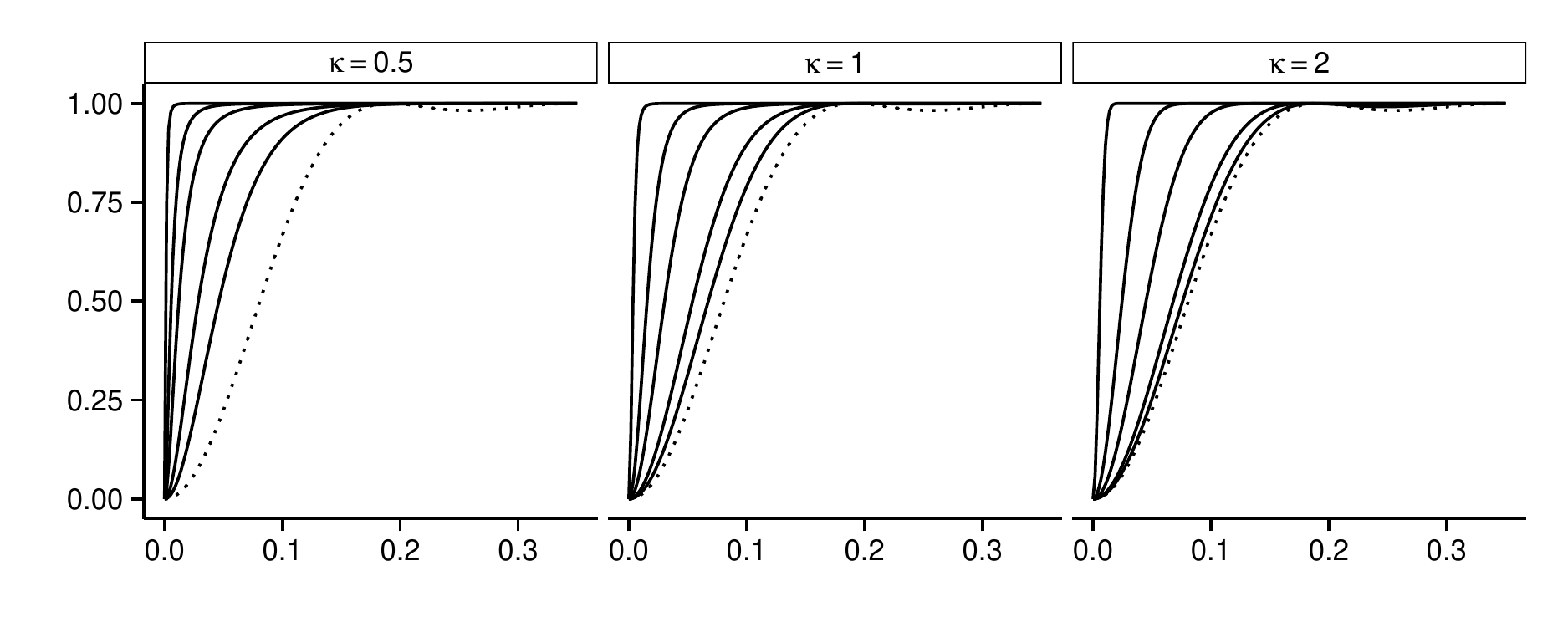}
  \caption{ Approximate pair correlation functions for DPP models with
    $d=2$.  Within each panel the full lines from left to right
    correspond to spectral models with $\beta=100,\allowbreak
    5,\allowbreak 1,\allowbreak 0.1,\allowbreak 0.01$ and $\alpha$
    chosen such that $\eta\approx400$ given the value of $\kappa$ as
    indicated in the figure.  The dotted line corresponds to the most
    repulsive DPP with $\eta=400$.  }
  \label{fig:specmodelpcf}
\end{figure}

The model covers a wide range of repulsiveness as indicated in
Figure~\ref{fig:specmodelpcf}.  Compared to the multiquadric model in
Figure~\ref{fig:MQpcf} we immediately notice that even with a moderate
value of the exponent parameter ($\kappa=2$) this model allows us to
come much closer to the most repulsive DPP.  The drawback of this
model is that neither the intensity nor the pair correlation function
is know analytically.  To approximate each pair correlation function
we have used \eqref{e:K0} with
$\beta_{\ell,d}=\lambda_{\ell,d}m_{\ell,d}/\eta$ where $\eta$ is
evaluated numerically.

\subsubsection{Mat\'{e}rn covariance functions}\label{s:Matern}

The Mat\'{e}rn correlation function is of class $\Psi_\infty^+$ and
given by
\begin{equation*}
  \psi(s)=\frac{2^{1-\nu}}{\Gamma(\nu)}\left(\frac{s}{c}\right)^\nu 
  K_\nu\left(\frac{s}{c}\right),\qquad0\le s\le\pi,
\end{equation*}
where $\nu\in(0,\frac{1}{2}]$ and $c>0$ are parameters and $K_\nu$
denotes the modified Bessel function of the second kind of order
$\nu$, see \cite[Section~4.5]{Gneiting:2013}. For $\nu=\frac{1}{2}$,
$\psi(s)=\exp(-s/c)$ is the exponential correlation function.
The nomenclature {\em Mat{\'e}rn} function might be considered a bit ambitious here, since we are considerably restricting the parameter space for $\nu$.
Indeed, it is true that any value of $\nu>0$ can be used when replacing the great circle distance with the chordal distance, but the use of this alternative metric is not contemplated in the present work. 

Suppose $R_0=\psi$ is (the radial part of) the kernel for an isotropic
DPP. Then the DPP becomes more and more repulsive as the scale
parameter $c$ or the smoothness parameter $\nu$ increases, since $g_0$
then decreases. In the limit, as $c$ tends to 0, $g_0$ tends to the
pair correlation function for a Poisson process. It can also be
verified that $g_0$ increases as $\nu$ decreases.

For $\nu=\frac{1}{2}$, $g_0(s)=1-\exp(-2s/c)$, so $g_0'(0)=2/c>0$ and
hence the DPP is locally less repulsive than any other DPP with the
same intensity and such that the slope for the tangent line of its
pair correlation function at $s=0$ is at most~$2/c$. For
$\nu<\frac{1}{2}$, $g_0'(0)=\infty$. That is, a singularity shows up
at zero in the derivative of $R_0$, which follows from the asymptotic
expansions of $K_\nu$ given in \cite[Chapter 9]{Abramowitz}. Hence,
the DPP is locally less repulsive than any other DPP with the same
intensity and with a finite slope for the tangent line of its pair
correlation function at $s=0$. Consequently, the variance condition
\eqref{e:varcond} is violated for all $\nu\in (0,1/2]$.

We are able to derive a few more analytical results: For $d=1$ and
$\nu=\frac{1}{2}$, \eqref{e:four} yields
\begin{equation*}
  \beta_{0,1}=\frac{c}{\pi}\left\{1-\exp\left(-\frac{\pi}{c}\right)\right\},
\end{equation*}
and
\begin{equation*}
  \beta_{\ell,1}=\frac{2}{\pi}
  \left\{1+(-1)^{\ell+1}
    \exp\left(-\frac{\pi}{c}\right)\right\}\frac{c}{1+c^2\ell^2}\, ,
  \qquad \ell=1,2,\ldots
\end{equation*}
Hence
\begin{equation*}
    \eta\tsub{max}{1}=
  \frac{\pi}{c}\,\frac1{1-\exp\left(-\frac{\pi}{c}\right)}\, ,
\end{equation*}
which is a decreasing function of $c$, with range $(1,\infty)$.

Incidentally, \cite{Guinness:Fuentes:16} introduced what they call a
circular Mat\'{e}rn covariance function and which for $d=1$ has Mercer
coefficients
\begin{equation*}
  \lambda_{\ell,1}=\frac{\sigma^2}{\left(\alpha^2+\ell^2\right)^{\nu+1/2}},\qquad\ell=0,1,\ldots,
\end{equation*}
where $\sigma>0$, $\nu>0$, and $\alpha>0$ are parameters. Consider a
DPP with the circular Mat\'{e}rn covariance function as its kernel and
with $d=1$. This is well defined exactly when
$\sigma\le\alpha^{\nu+1/2}$. Then the mean number of points is
\begin{equation*}
  \eta=\sum_{\ell=-\infty}^\infty
  \frac{\sigma^2}{\left(\alpha^2+\ell^2\right)^{\nu+1/2}},
\end{equation*}
which is bounded by
\begin{equation*}
    \eta\tsub{max}{1}= \sum_{\ell=-\infty}^\infty
  \frac{1}{\left(1+(\ell/\alpha)^2\right)^{\nu+1/2}}.
\end{equation*}
When $\nu$ is a half-integer, the kernel and hence also $\eta$ and
$\eta\tsub{max}{1}$ are expressible on closed form, see
\cite{Guinness:Fuentes:16}. Finally, considering the case
$\sigma=\alpha^{\nu+1/2}$ (equivalently $\eta=\eta\tsub{max}{1}$), then
\begin{equation*}
  \lambda_{\ell,1}=
  \frac{1}{\left(1+(\ell/\alpha)^2\right)^{\nu+1/2}},
  \qquad\ell=0,1,\ldots,
\end{equation*}
and we see that the DPP never reaches the most repulsive DPP except in
the limit where $\eta\tsub{max}{1}\rightarrow 1$.

\begin{figure}
  \centering
  \includegraphics[width=.4\textwidth]{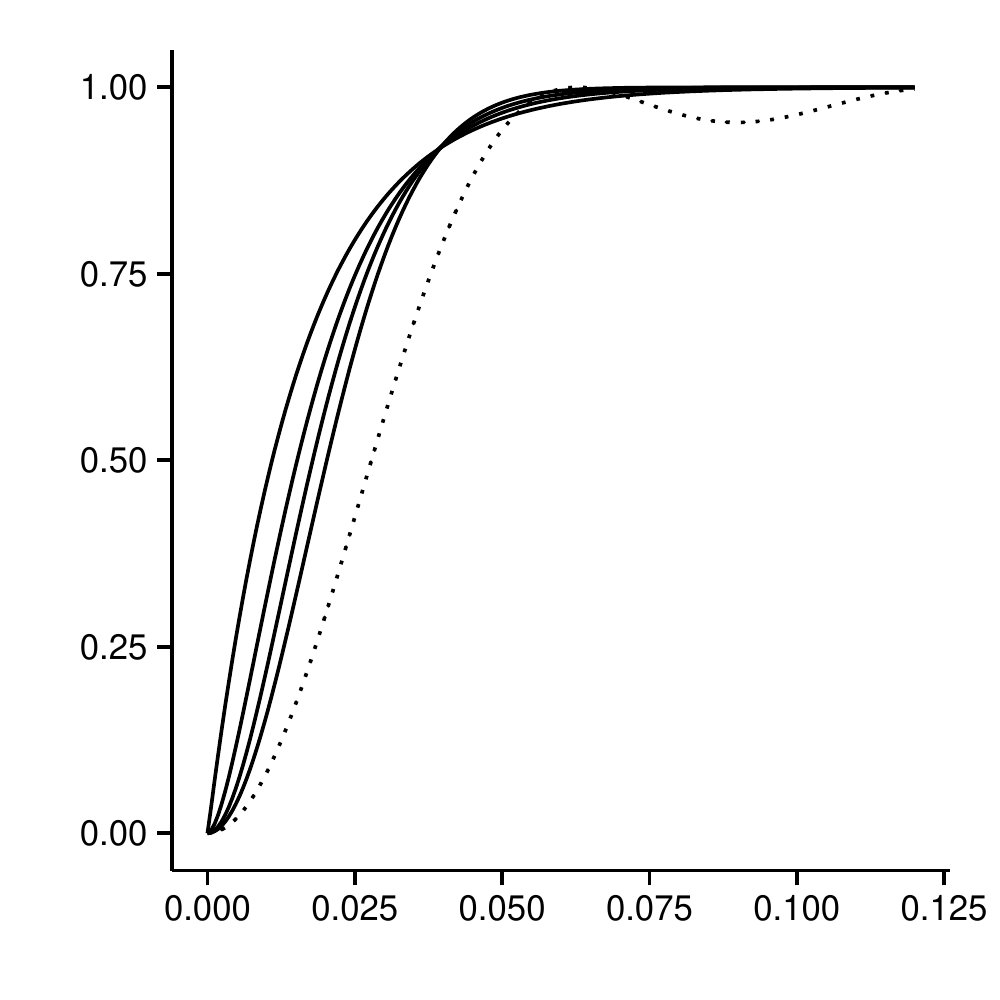}
  \caption{Pair correlation functions for DPP models with $d=1$.
    Full lines from left to right correspond to circular Mat\'{e}rn
    models with $\nu=0.5,1,2,10$, $\sigma=\alpha^{\nu+1/2}$, and
    $\alpha=31.8, 50, 75, 176.2$ chosen such that
    $\eta=\eta\tsub{max}{1}\approx100$.  The dotted line corresponds to the most
    repulsive DPP with $\eta=100$.  }
  \label{fig:circularmatern}
\end{figure}

Figure~\ref{fig:circularmatern} shows four different pair correlation
functions corresponding to different values of $\nu$ for the circular
Mat\'{e}rn model with $\eta=\eta\tsub{max}{1}$ (i.e. $\sigma=\alpha^{\nu+1/2}$).
For values $\nu>10$ the curves become almost indistinguishable from
the one with $\nu=10$, so we have omitted these from the figure.  The
left most curve ($\nu=1/2$) is in fact almost numerically identical to
the ordinary Mat\'{e}rn model introduced above with $\nu=1/2$ and
$\eta\tsub{max}{1}=100$.  Since $\nu=1/2$ corresponds to the most repulsive
ordinary Mat\'{e}rn model we see that the circular Mat\'{e}rn model
covers a much larger degree of repulsiveness than the ordinary
Mat\'{e}rn model.

\section{Concluding remarks}\label{s:final}

In this paper, we have considered determinantal point processes (DPPs)
on the $d$-dimensional unit sphere $\mathbb S^{d}$. We have shown that
DPPs on spheres share many properties with DPPs on $\mathbb R^d$, and
these properties are simpler to establish and to exploit for
statistical purposes on $\mathbb S^{d}$ due to compactness of the
space.

For DPPs with a distribution specified by a given kernel (a complex
covariance function of finite trace class and which is square
integrable), we have developed a suitable Mercer (spectral)
representation of the kernel. In particular, we have studied the case
of DPPs with a distribution specified by a continuous isotropic
kernel. Such kernels can be expressed by a Schoenberg representation
in terms of countable linear combination of Gegenbauer polynomials,
and a precise connection between the Schoenberg representation and the
Mercer representation has been presented. Furthermore, the trade-off
between the degree of repulsiveness and the expected number of points
in the model has been established, and the `most repulsive isotropic
DPPs' have been identified.
 
We have used the connection between the Schoenberg and Mercer
representations to construct a number of tractable and flexible
parametric models for isotropic DPPs.  We have considered two
different modelling approaches where we either work with a closed form
expression for the correlation function $\psi$ or with its
Mercer/spectral representation.  With the former approach the
multiquadric model seems to be the most promising model with some
flexibility, and the closed form expression for $\psi$ opens up for
computationally fast moment based parameter estimation in future work.
The two main drawbacks is that this class cannot cover the most
extreme cases of repulsion between points and simulation requires
truncation of a (possibly) infinite series, which in some cases may be
problematic.  However, in our experience the truncation works well in
the most interesting cases when we are not too close to a Poisson
point process.  In that case the Schoenberg coefficients decay very
slowly and it becomes computationally infeasible to accurately
approximate the $d$-Schoenberg coefficients.  The flexible spectral
model we have developed overcomes these two drawbacks.  It covers the
entire range of repulsiveness from the lack of repulsion in the
Poisson case to the most repulsive DPP, and it is straightforward to
generate simulated realizations from this model.  The main drawback
here is that the intensity and pair correlation function only can be
evaluated numerically making moment based inference more difficult,
and the parameters of the model may be harder to interpret.

We defer for another paper how to construct anisotropic models and to
perform statistical inference for spatial point pattern datasets on
the sphere.  In brief, smooth transformations and independent
thinnings of DPPs results in new DPPs, whereby anisotropic DPPs can be
constructed from isotropic DPPs. For $d=2$ particular forms for
anisotropy should also be investigated such as axial symmetry (see
e.g.\ \cite{jones:63,hitczenko:stein:12}) meaning that the kernel is
invariant to shifts in the polar longitude.

We have developed software in the \texttt{R} language \cite{R-core:15}
to handle DPPs on the sphere as an extension to the \texttt{spatstat}
package \cite{Baddeley:etal:15} and it will be released in a future
version of \texttt{spatstat}.  Until official release in
\texttt{spatstat} the code can be obtained by sending an email to the
authors. Some figures have been created with the \texttt{R} package
\texttt{ggplot2} \cite{Wickham:09}.

\subsection*{Acknowledgments}
Supported by the Danish Council for Independent Research | Natural
Sciences, grant 12-124675, ``Mathematical and Statistical Analysis of
Spatial Data'', by Proyecto Fondecyt Regular 1130647 from the Chilean Ministry
of Education, and by the Centre for Stochastic Geometry and Advanced
Bioimaging, funded by a grant (8721) from the Villum Foundation.

\appendix

\section{Simulation algorithm}\label{app:simulation}
As mentioned in the comment to Theorem~\ref{thm:dpp}\ref{item:simulation} we need to simulate points sequentially given the previously generated points.
This is done by rejection sampling from a uniform instrumental distribution as described in \cite{LMR2,LMR1} which requires an upper bound on the squared modulus of the eigenfunctions.
Specifically, when $d=2$ (using the notation from equation \eqref{eq:s_harmonics} in Theorem~\ref{thm:constlambda1}) we can use the bound
\begin{equation*}
     | Y_{l,k,2}(\vartheta, \varphi) |^2 \le \frac{2l+1}{4\pi} \frac{(l-|k|)!}{(l+|k|)!} \quad \text{for all } (\vartheta,\varphi)\in[0,\pi]\times[0,2\pi).
\end{equation*}

\section{Eigenvalues for nonnegative $\psi(s)$}\label{app:infimum}
Suppose that $\psi(s)$ is nonnegative. Then $\tilde{\psi}(x):=\psi(\arccos x)$ is nonnegative and according to \eqref{e:K0}
 \begin{equation*}
      \tilde{\psi}(x) =\sum_{\ell=0}^{\infty} \beta_{\ell,d} \frac{
        {\cal C}_{\ell}^{(\frac{d-1}{2})}(x)}{{\cal
          C}_{\ell}^{(\frac{d-1}{2})}(1)}, \qquad -1\le x\le 1.
    \end{equation*}
We use the orthogonality of the Gegenbauer polynomials with respect to the measure  $(1-x^2)^{\frac{d-2}2} \,\mathrm d x$ on $[-1,1]$, see \cite[p.\ 774]{Abramowitz},  to obtain 
\[\beta_{\ell,d}={\cal C}_{\ell}^{(\frac{d-1}{2})}(1)\cdot\frac{l!\, \left( 2l+d-1 \right)\left[\Gamma \left( \frac{d}2-\frac{1}2
 \right)\right]^{2}}{\pi \,{2}^{3-d}\Gamma  \left( l+d-1 \right) }
\int_{-1}^1 \tilde{\psi}(x)  {\cal C}_{\ell}^{(\frac{d-1}{2})}(x)(1-x^2)^{\frac{d-2}{2}} \,\mathrm d x.\]
It follows that
\begin{align}
\alpha_{\ell,d}&=\frac{\sigma_d\cdot\beta_{\ell,d}}{m_{\ell,d}}\nonumber\\
&=\sigma_d\cdot
\frac{[{\cal C}_{\ell}^{(\frac{d-1}{2})}(1)]^2}{m_{\ell,d}}
\cdot\frac {l!\, \left( 2l+d-1 \right)   \left[\Gamma \left( \frac{d}2-\frac{1}2\right)\right]^{2}}{\pi \,{2}^{3-d}\Gamma  \left( l+d-1 \right) }
\int_{-1}^1 \tilde{\psi}(x)  \frac{
        {\cal C}_{\ell}^{(\frac{d-1}{2})}(x)}{{\cal
          C}_{\ell}^{(\frac{d-1}{2})}(1)} (1-x^2)^{\frac{d-2}{2}} \,\mathrm d x\nonumber\\
          &=
          {\frac {2{\pi }^{d/2}}{\Gamma  \left( d/2 \right) }}
\int_{-1}^1 \tilde{\psi}(x)  \frac{
        {\cal C}_{\ell}^{(\frac{d-1}{2})}(x)}{{\cal
          C}_{\ell}^{(\frac{d-1}{2})}(1)} (1-x^2)^{\frac{d-2}{2}} \,\mathrm d x.\label{eq:alpha}
\end{align}
It is known that $\|{\cal C}_{\ell}^{(\frac{d-1}{2})}(x)/{{\cal
          C}_{\ell}^{(\frac{d-1}{2})}(1)}\|_{L^\infty}\leq 1$ for all $\ell\geq 0$,  see \cite[Theorem 7.32.1]{MR0372517}, and  ${\cal C}_{0}^{(\frac{d-1}{2})}(x)\equiv 1$. 
Combined with the fact that $\tilde{\psi}(x)(1-x^2)^{\frac{d-2}{2}}$ is nonnegative on $[-1,1]$, it follows directly from H\"older's inequality applied to \eqref{eq:alpha} that 
\[
\sup_\ell \alpha_{\ell,d}=\alpha_{0,d}=
 {\frac {2{\pi }^{d/2}}{\Gamma  \left( d/2 \right) }}
\int_{-1}^1 \tilde{\psi}(x) (1-x^2)^{\frac{d-2}{2}} \,\mathrm d x.\]
In fact, we may also conclude that $\alpha_{0,d}>\alpha_{\ell,d}$, $\ell\geq 1$, since  $|{\cal C}_{\ell}^{(\frac{d-1}{2})}(x)/{{\cal
         C}_{\ell}^{(\frac{d-1}{2})}(1)}|<1$ on a set of positive measure, which  can be deduced from the orthogonality of the Gegenbauer system.

\section{Proof of Theorem~\ref{thm:constlambda2}(b)}\label{app:proofconstlambda2}

For $d\geq 2$, we use \cite[Equation (2.7)]{MR3017365} to obtain
\begin{equation*}
  x^\ell=\sum_{\substack{n=0  \\ n-\ell\equiv 0\!\!\!\pmod 2}}^\ell\frac{(2n+d-1)(\ell!)\Gamma(\frac{d-1}2)}{2^{\ell+1} \{(\frac{\ell-n}2)!\}\Gamma(\frac{\ell+n+d+1}2)} \mathcal{C}_n^{(\frac{d-1}2)}(x),\qquad x\in [-1,1].
\end{equation*}
Hence, for any function $\psi$ admitting the expansion (\ref{e:K02})
we obtain an associated Schoenberg expansion of the type
\begin{align*}
  \psi(s)&=\sum_{\ell=0}^\infty \beta_{\ell}\Bigl( \sum_{
    \substack{n=0 \\ n-\ell\equiv 0\!\!\!\pmod 2}}^\ell
  \frac{(2n+d-1)(\ell!)\Gamma(\frac{d-1}2)}{2^{\ell+1}
    \{(\frac{\ell-n}2)!\}\Gamma(\frac{\ell+n+d+1}2)}
  \mathcal{C}_n^{(\frac{d-1}2)}(\cos s)
  \Bigr)\nonumber\\
  &=\sum_{n=0}^\infty \Bigl(\sum_{\substack{\ell=n\\n-\ell\equiv
      0\!\!\!\pmod 2}}^\infty \beta_{\ell}
  \frac{(2n+d-1)(\ell!)\Gamma(\frac{d-1}2)}{2^{\ell+1}
    \{(\frac{\ell-n}2)!\}\Gamma(\frac{\ell+n+d+1}2)}
  \mathcal{C}_{n}^{{(\frac{d-1}2)}}(1)\Bigr)
  \frac{\mathcal{C}_{n}^{(\frac{d-1}2)}(\cos
    s)}{\mathcal{C}_{n}^{{(\frac{d-1}2)}}(1)},
  \label{eq:Lexpansion}
\end{align*}
whereby \eqref{eq:coefficients} is seen to be true for $d\geq 2$,
since $\mathcal{C}_{n}^{{(\frac{d-1}2)}}(1)=\binom{n+d-2}{n}$.  For
$d=1$ we use a similar argument based on the power reduction formula,
\begin{align*}
  \cos^\ell s&=2^{-\ell}\sum_{ \substack{n=0 \\ n-\ell\equiv
      0\!\!\!\pmod 2}}^\ell
  \big(2-\delta_{n,0}\delta_{\ell\!\!\!\!\pmod 2,0}\big) \binom{\ell}{\frac{\ell-n}2} \cos(ns)\\
  &=2^{-\ell}\sum_{ \substack{n=0 \\ n-\ell\equiv 0\!\!\!\pmod
      2}}^\ell \big(2-\delta_{n,0}\delta_{\ell\!\!\!\!\pmod 2,0}\big)
  \binom{\ell}{\frac{\ell-n}2}\mathcal{C}_{n}^{(0)}(\cos s).
\end{align*}

\section{Proof of Theorem~\ref{prop:mostrep}}\label{app:proofmostrep}

This appendix verifies (a)--(c) in Theorem~\ref{prop:mostrep}.

(a) It follows straightforwardly from
\eqref{e:defrhon}--\eqref{e:defdpp} that
\begin{equation*}\label{e:starkk}
  I(g_0)=\frac{1}{\eta}-\frac{\text{Var}(\#\bX)}{\eta^2}.
\end{equation*} 
Considering independent Bernoulli variables $B_{\ell,k,d}$ with
parameters $\lambda_{\ell,d}$ for $k\in\mathcal K_{\ell,d}$ and
$\ell=0,1,\ldots$, cf.\ Theorem~\ref{thm:dpp}(b) and \eqref{e:nnn}, we
obtain
\begin{equation*}
  \text{Var}(\#\bX)=\sum_{\ell=0}^\infty
  m_{\ell,d}\lambda_{\ell,d}\left(1-\lambda_{\ell,d}\right).
\end{equation*}
Thereby \eqref{e:Ig0} follows. 

(b)--(c) The case $d=1$ is left for the reader. Now suppose
$d\ge2$. Consider $g_0=1-R_0^2$, with Schoenberg representation
\begin{equation}\label{eq:inter}
  R_0(s)=\sum_{\ell=0}^\infty\beta_{\ell,d}
  \frac{\mathcal C_\ell^{(\frac{d-1}{2})}(\cos s)}{\mathcal
    C_\ell^{(\frac{d-1}{2})}(1)}.
\end{equation}
We have $R_0(0)=1$,
\begin{equation*}
  \frac{\mathrm d\mathcal C_\ell^{(\frac{d-1}{2})}(\cos s)}{\mathrm ds}=
  -(d-1)\sin s\, \mathcal C_{\ell-1}^{(\frac{d+1}{2})}(\cos s)\,,
\end{equation*}
and
\begin{equation*}
  \bigg|\frac{(d-1)\sin s\, \mathcal C_{\ell-1}^{(\frac{d+1}{2})}(\cos
    s)}
  {\mathcal C_{\ell}^{(\frac{d-1}{2})}(1)} \bigg|\le
  (d-1)\frac{\binom{\ell+d-1}{\ell-1}}{\binom{\ell+d-2}{\ell}}
  =\frac{\ell(\ell+d-1)}{d},
\end{equation*}
see \cite[Chapter 17]{MR0393590}.  From this and the variance
condition \eqref{e:varcond} we deduce that termwise differentiation of
the right-hand side of \eqref{eq:inter} yields $R_0'(s)$.
In particular, $R_0'(0)=0$, and so $g'_0(0)=0$.  Further, for $s=0$,
\begin{equation*}
  \left[\frac{\mathrm d^2\mathcal C_\ell^{(\frac{d-1}{2})}(\cos s)}{\mathrm
      ds^2}\right]_{s=0}=
  \left[\ldots
    -(d-1)\cos s \,\mathcal C_{\ell-1}^{(\frac{d+1}{2})}(\cos s)\right]_{s=0}=
  -(d-1)\mathcal C_{\ell-1}^{(\frac{d+1}{2})}(1),
\end{equation*}
where `$\ldots$' is a product involving the $\sin^2s$ term and hence
is 0 at $s=0$.  Therefore, using again \eqref{e:varcond} and the same
justification of termwise differentiation as above,
\begin{equation*}
  g_0''(0)=-2R_0''(0)=2(d-1)\sum_{\ell=0}^\infty\beta_{\ell,d}
  \frac{\binom{\ell+d-1}{\ell-1}}{\binom{\ell+d-2}{\ell}}
  =\frac{2}{d}\sum_{\ell=0}^\infty\ell(\ell+d-1)\beta_{\ell,d}.
\end{equation*}
Thereby \eqref{e:g0kappa} is verified. Finally,
\begin{equation*}
  g_0''(0)=\frac{2}{d\eta}\sum_{\ell=1}^\infty
  \ell(\ell+d-1)m_{\ell,d}\lambda_{\ell,d}
  \, ,
\end{equation*}
and since $\ell(\ell+d-1)m_{\ell,d},\ \ell=1,2,\ldots,$ is a strictly
increasing sequence, we conclude that $\bX^{(\eta)}$ is the unique
locally most repulsive isotropic DPP.

\section{Askey and Wendland correlation functions}\label{app:Askey}

Consider the Askey function
\begin{equation*} \label{askey} \psi_{\nu}(t) = \left ( 1- t\right
  )_+^{\nu} , \qquad t\ge0,\qquad \nu>0.
\end{equation*} 
Denote $\widetilde{\psi}_{\nu}$ the restriction of $\psi_{\nu}$ to the
interval $[0,\pi]$. In fact $\psi_{\nu}$ is the radial part of an
Euclidean isotropic correlation function defined on $\mathbb R^{d}$ if
and only if $\nu \ge \frac{d+1}{2}$, cf.\ \cite{Zastavnyi:Trigub:02}.
Gneiting \cite{Gneiting:02} in his {\em essay} used instead the
condition $\nu \ge \lfloor d/2+1\rfloor$. This means that the function
$\varphi(\bx)=\psi_{\nu}(\norm{\bx})$ for
$\bx\in\mathbb R^d$ is positive definite under the mentioned
constraint on $\nu$. Additionally, $\varphi$ is compactly supported on
the unit ball of $\mathbb R^d$, and it can be arbitrarily rescaled to
any ball of $\mathbb R^d$ with radius $c>0$ by considering
$\varphi(\cdot/c)$. Thus $\widetilde{\psi}_{\nu} \in \Psi_3^+$
provided $\nu \ge 2$, cf.\ \cite[Theorem~3]{Gneiting:02}.

Wendland functions are obtained through application of the Mont{\'e}e
operator \cite{matheron:73}
to Askey functions. 
The Mont{\'e}e operator is defined by 
\begin{equation*}
  {\cal I} \psi_{\nu}(t)= \frac{\int_{t}^{\infty} u \psi_{\nu} \left ( \frac{u}{b} \right ) {\rm d} u }{\int_{0}^{\infty} u \psi_{\nu} \left ( \frac{u}{b} \right ) {\rm d} u }, \qquad t >0,
\end{equation*}
and we 
define for $k=1,2,\ldots$, $\psi_{\nu,k}$ as the $k$th iterated
application of the Mont{\'e}e operator to $\psi_{\nu}$, and we set
$\psi_{\nu,0}=\psi_{\nu}$. Arguments in Gneiting (2002) show that
$\varphi(\boldsymbol{x})= \psi_{\nu,k}(\|\boldsymbol{x}\|)$,
$\boldsymbol{x} \in \mathbb R^d$, is positive definite provided $\nu
\ge \frac{d+1}{2}+k$. Thus, for $k=0,1,\ldots$, the restriction of
$\psi_{\nu,k}$ to $[0,\pi]$, denoted 
$\widetilde{\psi}_{\nu,k}$, belongs to $\Psi_{3}^{+}$ provided $\nu
\ge 2 +k $.

We shall only work with the special case $\nu= \lfloor d/2 + k
+1\rfloor$, 
where it is possible to deduce a closed form for the associated
spectral density, see \cite{Zastavnyi:06}.
We consider then the mapping 
$\widetilde{\psi}_{\nu,k}$ as the restriction to $[0,\pi]$ of
$\psi_{\nu,k}$, $k=0,1,\ldots$.
Note that for $d=1$, 
the one-Schoenberg coefficient for $\psi_{\nu,k}$ can be calculated
straightforwardly from the Fourier transform \eqref{e:four} using
partial integration.  Thereby we obtain Table~\ref{table1}, where
$\widetilde{\psi}_{\nu,0}$ is the Askey function,
$\widetilde{\psi}_{\nu,1}$ is the $C^2$-Wendland function, and
$\widetilde{\psi}_{\nu,2}$ is the $C^4$-Wendland function.

\bibliographystyle{abbrv} \bibliography{harmonic}
\end{document}